\documentclass[10pt]{IEEEtran}
\IEEEoverridecommandlockouts

\usepackage[noadjust]{cite}
\usepackage{amsmath}
\usepackage{amssymb}
\usepackage{amsthm}
\usepackage{mathtools}
\usepackage{bbm}
\usepackage{algorithm}
\usepackage[noend]{algpseudocode}
\usepackage{bm}
\usepackage{comment}
\usepackage{hyperref}
\hypersetup{
    colorlinks,
    linkcolor={black},
    citecolor={black},
    urlcolor={blue!80!black}
}
\usepackage{tabularx,booktabs,multirow, multicol, makecell, adjustbox}
\usepackage[table,xcdraw]{xcolor}
\usepackage[caption=false]{subfig}
\usepackage[capitalize]{cleveref}
\usepackage[normalem]{ulem}
\usepackage{pgfplots}
\pgfplotsset{
    compat=1.3,
    legend style={font=\scriptsize, fill opacity=1,  draw opacity=1, text opacity=1, draw=white!15!black, legend cell align=left, align=left},     
    ymajorgrids=true,
    xmajorgrids=true,    
    yminorticks=false,
    xminorticks=false,
    grid style={dashed},
    title style={font=\small},
    label style={font=\footnotesize},
    tick label style={font=\footnotesize},    
    tick align=inside,
    axis background/.style={fill=white},
    ylabel shift=-3pt,
}
\usepgfplotslibrary{fillbetween}
\usetikzlibrary{patterns,arrows,plotmarks}
\usepgfplotslibrary{groupplots}
\pgfdeclarelayer{background}
\pgfsetlayers{background,main}
\usetikzlibrary{automata,positioning}
\usetikzlibrary{decorations}
\usetikzlibrary{shapes.arrows}
\usetikzlibrary{tikzmark}
\usetikzlibrary{calc}
\usetikzlibrary{decorations.markings}
\algrenewcommand\algorithmicindent{10pt}
\usepgfplotslibrary{colorbrewer}

\newtheorem{approximation}{Approximation}

\newtheorem{example}{Example}

\newtheorem{scenario}{Scenario}

\newcommand{\T}{^{\intercal}}     
\newcommand{\E}[1]{\mathbb{E}\left[ #1 \right]} 
\newcommand{\mc}[1]{\mathcal{#1}}   
\newcommand{\mb}[1]{\mathbf{#1}}    
\DeclareMathOperator*{\argmax}{arg\,max}    

\usepackage{url}

\usepackage[acronym]{glossaries}
\glsdisablehyper

\newacronym{3gpp}{3GPP}{3rd Generation Partnership Project}

\newacronym{afsa}{AFSA}{adaptive framed slotted ALOHA}
\newacronym{aoi}{AoI}{age of information}
\newacronym{aoii}{AoII}{age of incorrect information}

\newacronym{bs}{BS}{base station}

\newacronym{cdf}{CDF}{cumulative distribution function}
\newacronym{cfc}{CFC-push/pull}{contention-free pull and contention-push}
\newacronym{cra}{CRA}{cluster risk aware}

\newacronym{dl}{DL}{downlink}
\newacronym{dt}{DT}{digital twin}

\newacronym{fsa}{FSA}{framed slotted ALOHA}

\newacronym{gnb}{gNB}{next generation Node B}

\newacronym{hmm}{HMM}{Hidden Markov model}

\newacronym{iid}{i.i.d.}{independent and identically distributed}
\newacronym{iot}{IoT}{Internet of Things}

\newacronym{map}{MAP}{maximum a posteriori}
\newacronym{mac}{MAC}{medium access control}
\newacronym{maf}{MAF}{maximum age first}
\newacronym{mc}{MC}{Monte Carlo}
\newacronym{mse}{MSE}{mean squared error}

\newacronym{ofdm}{OFDM}{orthogonal frequency-division multiplexing}
\newacronym{pdf}{PDF}{probability density function}
\newacronym{phy}{PHY}{physical}
\newacronym{pmf}{PMF}{probability mass function}
\newacronym{pps}{PPS}{push-pull scheduler}
\newacronym{prb}{PRB}{physical resource block}

\newacronym{qaoi}{QAoI}{query age of information}

\newacronym{re}{RE}{resource element}
\newacronym{rsm}{RSM}{reactive subframe manager}

\newacronym{sd}{SD}{standard deviation}
\newacronym{sic}{SIC}{successive interference cancellation}
\newacronym{ss}{SS}{synchronization signal}
\newacronym{ssm}{SSM}{stable subframe manager}

\newacronym{ul}{UL}{uplink}
\newacronym{wsn}{WSN}{wireless sensor network}

\definecolor{color0}{HTML}{00429D}
\definecolor{color1}{HTML}{844D99}
\definecolor{color2}{HTML}{C3608E}
\definecolor{color3}{HTML}{EF8078}
\definecolor{color4}{HTML}{FFB047}
\definecolor{amaranth}{rgb}{0.9, 0.17, 0.31}

\pgfplotstableset{col sep=comma}

\def \fwidth{\columnwidth}

\def \fheight {0.4\columnwidth}

\definecolor{lightgray}{HTML}{999999}
\definecolor{color0}{HTML}{00429D}
\definecolor{color1}{HTML}{915a8f}
\definecolor{color2}{HTML}{C3608E}
\definecolor{color3}{HTML}{d27f76}
\definecolor{color4}{HTML}{FFB047}

\begin{document}
\title{Push-Pull Medium Access for Digital Twin Alignment and Low-Latency Anomaly Reporting}

\author{Federico Chiariotti,~\IEEEmembership{Senior Member,~IEEE,} Fabio Saggese,~\IEEEmembership{Member,~IEEE,} Andrea Munari,~\IEEEmembership{Senior Member,~IEEE,} Leonardo Badia,~\IEEEmembership{Senior Member,~IEEE,} and Petar Popovski,~\IEEEmembership{Fellow,~IEEE}\thanks{Federico Chiariotti and Fabio Saggese contributed equally to this work. F. Chiariotti (federico.chiariotti@unipd.it) and L. Badia (leonardo.badia@unipd.it) are with the Dept. of Information Engineering, University of Padova, Padua, Italy. F. Saggese (fabio.saggese@ing.unipi.it) is with the Dept. of Information Engineering, University of Pisa, Italy. A. Munari (andrea.munari@dlr.de) is with the Inst. of Communications and Navigation, German Aerospace Center (DLR), We{\ss}ling, Germany. P. Popovski (petarp@es.aau.dk) is with the Dept. of Electronic Systems, Aalborg University, Aalborg \O{}st, Denmark. This work was supported, in part, by the Velux Foundation, Denmark, through the Villum Investigator Grant WATER, nr. 37793, and partly by the Horizon Europe SNS projects ``6G-GOALS'' (grant nr. 101139232) and ``MAGIC-6G'' (grant nr. 101292933). F. Saggese's work is funded by Horizon Europe MSCA Postdoctoral Fellowships with Grant~101204088. A. Munari would like to thank the Federal Ministry of Research, Technology, and Space (BMFTR) for supporting the xG‑RIC project as part of the research program Communication Systems  ``Souverän. Digital. Vernetzt.'' (grant number 16KIS2429K).}}

\maketitle

\begin{abstract}
A \gls{dt} contains a set of virtual models of real systems and processes that are synchronized with their physical counterparts. 
In a setup in which contact with the physical world is maintained through sensors and actuators that are wirelessly connected to the \gls{dt}'s computing engine, \gls{dt} alignment requires periodic status updates, while safety‑critical messages and fault conditions call for low‑latency anomaly reporting, creating a fundamental trade‑off in how wireless resources are used. 
We present a medium access framework combining \emph{pull-based} updates, centrally scheduled according to goal-oriented principles, with urgent \emph{push-based} updates, for which transmission decisions are made directly by the sensors. This enables the system to quickly detect and recover from anomalies while maintaining \gls{dt} alignment. We thus design a \gls{pps} that strikes a balance in the trade-off between \gls{dt} alignment in normal conditions and anomaly reporting, optimizing resource usage and reducing \gls{dt} drift by $20-30\%$ with respect to state-of-the-art solutions while maintaining the same anomaly detection guarantees, or reducing worst-case anomaly detection times by $30-70\%$ while meeting the same \gls{dt} alignment conditions.
\end{abstract}

\glsresetall

\section{Introduction}

The \gls{dt} paradigm has emerged as a transformative approach to bridge the gap between physical systems and their virtual representations~\cite{Wright2020dt}. An application \gls{dt} includes a comprehensive set of virtual models that mirror real systems and processes, maintaining continuous synchronization with their physical counterparts through sensor data streams.
It enables real-time monitoring, prediction, and control across diverse industrial applications~\cite{qin2024machine}, along with accelerated experimentation and examination of counterfactuals in several scenarios. Additionally, \glspl{dt} have co-evolved with 5G systems and Edge computing~\cite{yang2024joint}, enabling modular platforms~\cite{duran2026toward} that can run and combine multiple Twins, control synchronization latency, and allocate computing and communication resources~\cite{yu2025optimizing}.

The utility of \glspl{dt} relies on maintaining precise alignment between virtual models and the evolving physical reality they represent. This alignment is highly dependent on the specific system, as well as the purpose the \gls{dt} is being used for, so there are several metrics for \gls{dt} alignment~\cite{ak2026ai}, but timing limits are often used as a general proxy~\cite{cakir2025time}: if the latest update from the physical system is recent, the \gls{dt} will be aligned with a high probability. 
However, timing requirements are inflexible, and do not consider the content and the importance of different updates. The degree of \gls{dt} misalignment can be measured directly through a \emph{\gls{dt} drift} metric, and the allocation of the wireless resources can be tailored to the dynamics of this drift.
This involves a deeper integration between the \gls{dt} and the communication system, as on one side, communication decisions are optimized in a \emph{goal-oriented} fashion to minimize the \gls{dt} drift, while on the other, the \gls{dt} can exploit both explicit and implicit information from the communication system~\cite{Chiariotti2025delta} to improve its estimates.

This joint optimization involves a key trade-off, as it requires balancing \emph{proactive} system monitoring of gradual state changes, i.e., polling of nodes to maintain synchronization or minimize the drift metric, with \emph{reactive} responses to critical anomalies~\cite{moyne2020requirements}, which must come from the sensors themselves, as anomalies are hard to predict by their own nature. Although both types of updates are essential for trustworthy \gls{dt} operations, they impose conflicting demands on the network.

Proactive monitoring typically involves fusing data from multiple sensors to detect system drift or degraded working states~\cite{rivera2022designing}. Although individual sensor readings may not show strong deviations, their combined patterns can reveal subtle drift, requiring model updates. This reflects the predictive role of \glspl{dt}: the system periodically requests data from sensors to refine its understanding of the physical state, enabling timely detection of change, while still tolerating moderate delays.
In contrast, reactive alerting addresses urgent events that require immediate attention~\cite{castellani2020real,de2023digital}, such as equipment faults or safety violations. Here, individual sensors can independently detect and report outlier values, guaranteeing timely and reliable responsiveness in worst-case conditions.

\begin{example} \label{example}
Consider an industry with sensors that measure machine temperature and vibration. The \gls{dt} system benefits by regularly aggregating sensor data to detect gradual degradation patterns that indicate maintenance needs. However, when a bearing begins to fail, triggering sudden measure spikes, individual sensors must immediately alert the system without waiting for the next collection cycle. Consider $M = 3$ machines, each monitored by $4$ sensors, which can jointly detect drifts in machine behavior or individually sense rare anomalies. Since not all sensors must report to identify model drift, assume detection succeeds with probability $\nu_\mathrm{reset} = 90\%$ if $3$ out of $4$ sensors respond.
Suppose communication occurs over frames with $4$ slots. A pull-only strategy could cycle through sensor groups, yielding an average detection delay of $(M-1)/2 = 1$ frame, which is acceptable for drift but unsuitable for anomalies. 
A better solution pulls data from only $3$ sensors in the first three slots, reserving the fourth as a ``push opportunity'' for anomaly reporting. This increases drift detection delay to $(M-1)/2 + M(1-\nu_\mathrm{reset})/\nu_\mathrm{reset}\simeq 1.3$ frames, but enables immediate anomaly reporting, assuming no collisions. The scheme remains effective if anomalies occur less than once every $32$ frames. All incoming messages, whether push or pull, contribute to updating the \gls{dt}, enhancing system performance beyond either mode alone.
\end{example}

Despite the benefits of combining the pull and push paradigms, several technical challenges have limited the development of adaptive unified frameworks~\cite{Agheli2025pushpull, cavallero2024co-existence, Shiraishi2024pushpull}: \emph{(i}) designing frames that support both contention-free pull and contention-based push for multiple sensors while maintaining existing standard compatibility presents significant engineering challenges; \emph{(ii}) dynamically allocating resources between pull and push requires balancing competing objectives while adapting to changing system conditions; \emph{(iii}) maintaining accurate beliefs about drift and anomalies requires robust tracking handling partial observations, communication failures, and the interdependence between different information sources. 

We address these challenges by proposing a medium access framework that dynamically allocates push-pull resources through a novel \gls{pps}. 
Our framework employs a frame structure that can be easily integrated with \gls{3gpp} standards, enabling the \gls{bs} to continuously update the \gls{dt} by collecting observations from the sensors. The key contributions of this work are: \emph{(i)}, we formalize a unified mathematical model that captures both \gls{dt} drift and anomaly detection requirements, providing a general framework with two different application examples;
    \emph{(ii)}, we design intelligent scheduling algorithms that dynamically allocate resources to pull and push traffic based on real-time system state estimates;
    \emph{(iii)}, we derive a belief update system that can leverage implicit information from protocol actions to gain additional insights on potential anomalies; and \emph{(iv)}, we thoroughly evaluate the new framework by Monte Carlo simulation, considering state-of-the-art benchmarks and potential \gls{dt} model imperfections. 
    
The performance gains we obtain are significant, as \gls{pps} can reduce \gls{dt} drift by approximately $20-30\%$ or reduce the \gls{aoii} of unreported anomalies by $30-70\%$ while guaranteeing a limited \gls{dt} drift. A preliminary version of this work~\cite{chiariotti2026combined} presented a basic version of the framework, considering only a limited drift model without any resource management schemes between push and pull traffic.

The rest of this paper is organized as follows: first, we present the related work on protocols and metrics for \gls{dt} alignment and information freshness in Sec.~\ref{sec:related}.
The system model is described in Sec.~\ref{sec:system}, while the \gls{dt} updates and anomalies tracking mechanism is presented in Sec.~\ref{sec:tracking}. The scheduler is defined in Sec.~\ref{sec:scheduler}, and simulation results are discussed in Sec.~\ref{sec:results}. Finally, Sec.~\ref{sec:concl} concludes the paper.

\section{Related Work}\label{sec:related}

Freshness-related metrics introduce a key principle in networked systems related to timely reporting of relevant information. A foundational step towards the inclusion of decision-driven status updates and the separation between push- and pull-based communication is provided in~\cite{sun2017update}, which highlights how update decisions should be triggered by system state considerations, rather than individual pushes alone. However, this seminal work considers a push-based mechanism with a single source, whereas we argue that the requirements of a \gls{dt} representation are often nuanced, and a push-pull dichotomy among multiple sensors is naturally present.

Pull-based monitoring seems like a natural choice for \gls{dt} alignment, as it allows a central node to explicitly request data from sensors as dictated by the \gls{dt}. Thus, pull-based approaches looks particularly well-suited for periodic updates ``on demand,'' e.g., for goal-oriented control tasks or scenarios where the central system can optimally schedule transmissions based on global system knowledge.
For example,~\cite{Chiariotti2022query} captures the freshness of information specifically at the instant when it is needed through the \gls{qaoi}. This shifts the focus from continuous monitoring to event-driven access, revealing that optimal pull-based policies fundamentally differ from their push-based counterparts, as the timing of queries becomes the primary control variable. Further developments in~\cite{ildiz2023pull} investigate the trade-off between querying and waiting, showing that optimal policies exhibit a threshold structure that depend on both system dynamics and query costs. 
Similarly,~\cite{zakeri2025gopull} considers remote tracking with correlated observations in a pull-based setting, emphasizing that the structure of the underlying information source is key in determining optimal query strategies.
This reasoning can also be extended to more complex prediction models, such as deep learning-based conformal prediction~\cite{homaei2026conformal}.

On the other hand, push-based communication approaches allow sensors to autonomously transmit data when they detect significant changes or anomalous conditions, thus improving the reactivity of the system to these unexpected events, which must be reported as soon as possible. Some recent works have explored how adaptive push-based policies can improve freshness performance in wireless \gls{iot} systems~\cite{bedewy2017age}. More complex push-based systems based on random access protocols~\cite{israel2017ra,yue2023age} excel where event-driven communication is more efficient than periodic polling, especially for rare anomalous events that require immediate attention~\cite{munari2024monitoring,Chiariotti2025delta, Cosandal2025aoii}. 

While these approaches have often been viewed as alternative strategies~\cite{pandey2025pullpushmag}, recent works suggest that their combination can yield significant performance improvements. The creation of an underlying pull-based structure in which individual push-based updates can be inserted, based on random access protocols, extends the approach by enabling sensors to report anomalies, obtaining an intrinsic \emph{coexistence gain}~\cite{cavallero2024co-existence,Srivastava2025pushpull}. The coexistence of push and pull mechanisms can be investigated in wireless \gls{iot} systems, showing that combining proactive transmissions with on-demand queries  significantly improves performance under shared resource constraints, as reported in~\cite{Shiraishi2024pushpull} for the specific context of wake-up radio architectures. 
This line of work is further extended in~\cite{Agheli2025pushpull}, where a unified push-pull framework is proposed and analyzed under goal-oriented metrics, demonstrating that joint optimization of the two paradigms enables a more efficient use of communication resources. 

Overall, the literature shows a clear transition from periodic updates to more flexible, query-driven and hybrid communication methods, with an increasing emphasis on goal-oriented performance metrics and on the interplay between communication decisions and \gls{dt} models.  
However, the interplay between pull-based information access and decentralized decision-making remains only partially understood, particularly in settings where multiple agents independently decide when to request or generate updates. 
This gap motivates the development of models that jointly capture query-driven communication, system-level interactions, and the resulting impact on information freshness and efficiency.

\section{System Model}\label{sec:system}
Consider a \gls{wsn} in which a set $\mc{N}$ of sensors, with $|\mc{N}|=N$, observe a number of spatially distributed processes, mapped to their \gls{dt}, and communicate with a \gls{bs}. Our dual objective is to maintain the \gls{dt} of the system aligned with physical reality, as well as detecting anomalous conditions as quickly as possible. 

In the following, we consider two possible systems that are very different from each other, yet they both fit our general model of a \gls{dt}. The examples cover several practical \gls{iot} use cases, but the definition of the system model and the corresponding solutions have a wider applicability.

\begin{scenario}[\gls{hmm}]
\label{scen:hmm}
  In the \gls{hmm} scenario, the drift behavior of the \glspl{dt} is represented by a \emph{discrete Markov chain}. In this case, we can denote a set of \emph{drift states}, i.e., states in which the error of the \gls{dt} exceeds a tolerance threshold. The \gls{bs} also has a reset mechanism: whenever it becomes aware of the drift with a given confidence, it realigns the \gls{dt} with the new measurements. This mechanism can represent a variety of cases in which there is a strict tolerance on \gls{dt} drift, and discrete \glspl{hmm} are often used to implement \glspl{dt} of simple systems~\cite{ghosh2019hidden,paras2025reliability}.   
\end{scenario}

\begin{scenario}[Kalman]
\label{scen:kalman}
 In this scenario, the \gls{dt} is a Kalman filter~\cite{kalman1960new} running over a \emph{linear dynamical system}. In this case, the drift of the \gls{dt} is mapped to the \gls{mse} of the continuous state estimate, which can be derived through the Kalman filter equations. This is a common use case in tracking systems~\cite{castejon2025enhanced,donato2024self}, as the Kalman filter is a well-known and robust tool with several extensions to deal with non-linear or unknown systems and noise processes.   
\end{scenario}

\subsection{Drift Model}\label{ssec:system_drift}
We denote the set of clusters of sensors, each of which is mapped to its own \gls{dt}, as $\mc{D}=\{\mc{C}^{(1)},\ldots,\mc{C}^{(D)}\}$. We assume that $\cup_{i=1}^D \mc{C}^{(i)}=\mc{N}_d\subseteq\mc{N}$ and $\mc{C}^{(i)}\cap\mc{C}^{(j)}=\varnothing\, \forall i\neq j$, i.e., only a subset of the sensors might be part of a cluster, and clusters are mutually exclusive. We denote the number of nodes in cluster $i$ as $C_i=\left|\mc{C}^{(i)}\right|$.

Time is divided into \emph{frames} of equal duration, further detailed in Sec.~\ref{ssec:system_comms}.
We model the evolution of each cluster as a stochastic process, in which observations convey information about a hidden state, representing the drift between the physical system and its \gls{dt}. Individual sensors make observations that might be affected by errors and imperfections, which are used to estimate the hidden state for cluster $i$ at frame $k$, ${x}^{(i)}(k)\in\mc{X}^{(i)}$, where $\mc{X}^{(i)}$ denotes the set of possible states. We define a \emph{drift function} $e^{(i)}(x):\mc{X}^{(i)}\to\mathbb{R}^+$, which denotes the misalignment of the \gls{dt} with its physical counterpart. The \gls{bs} maintains the estimate over the state of the \gls{dt} and its drift with sensor observations, which can be used to make scheduling decisions.

 The transition probability function $\mc{F}^{(i)}:\mc{X}^{(i)}\to\mc{P}(\mc{X}^{(i)})$, where $\mc{P}(\mc{X}^{(i)})$ is the probability simplex over the state space, defines the state evolution of each cluster. While the \gls{bs} has a complete view of each component or environment through the \gls{dt}, individual sensors are unable to detect drift, as it may depend on complex patterns involving multiple components of the overall system, while they only have access to local measurements.

\subsubsection{HMM} In Scenario~\ref{scen:hmm}, we model the state space $\mc{X}^{(i)}$ as a discrete set, over which we define a set of drift states, $\mc{X}^{(i)}_d$. The drift function is $\mc{I}\big(x\in\mc{X}^{(i)}_d\big)$, where $\mc{I}(\cdot)$ is the indicator function, equal to $1$ if the Boolean expression inside the argument is true and $0$ otherwise. The state of the \gls{hmm} is reset if the estimated probability of being in a drift state is greater than $\nu_{\text{reset}}$, and the state update function $\mc{F}^{(i)}$ takes the form of a transition matrix $\mb{U}^{(i)}$ over the state space $\mc{X}^{(i)}$~\cite{chiariotti2026combined}.

\subsubsection{Kalman} In Scenario~\ref{scen:kalman}, the state space is $\mc{X}^{(i)}\subseteq\mathbb{R}^M$, where the dimension $M$ is defined by the system model. The state update equation for the system, defining the transition probability $\mc{F}^{(i)}$, is
\begin{equation}
\mb{x}^{(i)}(k+1)=\mb{F}^{(i)} \, \mb{x}^{(i)}(k)+\mb{w}(k).
\end{equation}
 where $\mb{F}^{(i)}\in\mathbb{R}^{M\times M}$ is a known matrix and $\mb{w}(k)\sim\mc{N}(\mb{0}_M,\bm{\Sigma}_w)$ is a Gaussian noise with covariance matrix $\mb{\Sigma}_v\in\mathbb{R}^{M\times M}$. The observation function is
 \begin{equation}
     \mb{o}^{(i)}(k)=\mb{H}^{(i)}\mb{x}^{(i)}(k)+\mb{v}^{(i)}(k),
 \end{equation}
 where $\mb{H}^{(i)}\in\mathbb{R}^{C_i\times M}$ is the observation matrix and $\mb{v}^{(i)}(k)\sim\mc{N}\left(\mb{0}_{C_i},\mb{\Sigma}_v\right)$ is a Gaussian observation noise with covariance matrix $\mb{\Sigma}_v\in\mathbb{R}^{C_i\times C_i}$.
The filter maintains an estimate $\hat{\mb{x}}^{(i)}(k)$ and a prediction covariance matrix $\mb{P}^{(i)}(k)$, and the drift function is the value of the \gls{mse}, $e(\mb{x},\hat{\mb{x}})=(\hat{\mb{x}}-\mb{x})^\intercal(\hat{\mb{x}}-\mb{x})$. 

\subsection{Anomaly Model}

Assume that anomalies involve a single sensor, i.e., they appear as abnormal measurements over a specific element of the process. We denote the subset of sensors that observe anomalies as $\mc{N}_a$, with $N_a=|\mc{N}_a|$.
We do not make any assumptions about the intersection of sets $\mc{N}_a$ and $\mc{N}_d$: arbitrary subsets of nodes may be part of a \gls{dt} model, observe anomalies, or both.\footnote{This flexibility is critical, as we would otherwise have to define anomalies as information updates not captured by the \gls{hmm}, which would require additional assumptions about side knowledge at the \gls{bs} and the sensors.} Naturally,  $\mc{N}_a\cup\mc{N}_d=\mc{N}$, as nodes that perform neither task can be removed.

We define the anomaly state vector as the stochastic process $\mb{y}(k)\in\{0,1\}^{N_a}$, where $y_n(k)=1$ indicates that sensor $n$ observes an anomaly in frame $k$. We model the appearance of anomalies for each sensor as an independent Markov chain, whose transition matrix $\mb{M}_n$ is
\begin{equation}
  \mb{M}_n=\begin{pmatrix}
      1-\lambda_n& \lambda_n\\
      \mu_n+\xi_{n}(k)-\mu_n\xi_{n}(k) & (1-\xi_n(k))(1-\mu_n)
  \end{pmatrix},
\end{equation}
where $\lambda_n$ is the anomaly rate in each frame and $\mu_n$ is the spontaneous resolution rate in each frame, i.e., the probability that the anomaly disappears without the need for external intervention. Finally, $\xi_n(k)$ is an indicator variable equal to $1$ if sensor $n$ successfully reports the anomaly during frame $k$, so that the system operator can solve the issue. Vectors $\bm{\lambda}\in\mathbb{R}_+^{N_a}$ and $\bm{\mu}\in\mathbb{R}_+^{N_a}$ collect each sensor's anomaly and resolution rates $\lambda_n$ and $\mu_n$, which are known to the \gls{bs}.

We define the \gls{aoii} of sensor $n$ after frame $k$, $\Theta_{n}(k)$, as the time that node $n$ has spent in state $1$~\cite{maatouk2023aoii}:
\begin{equation}
    \Theta_{n}(k)=k-\sup\left\{\ell\in\{0,\ldots,k\}:y_{n}(\ell)=0\right\}.
\end{equation}
Anomalies are immediately identifiable by the involved sensors, as the observed value can be directly compared with the normal operation range. As such, they are amenable to push-based approaches, in which each sensor decides when and whether to report them, based on its own observation of the anomaly and commands from the \gls{bs}. 

\subsection{Communication Model}\label{ssec:system_comms}

\begin{figure}[tb]
    \centering
    \includegraphics[width=0.99\columnwidth]{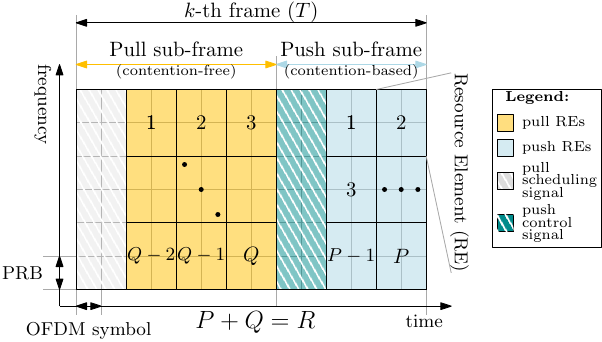}
    \caption{Visualization of the MAC frame translated to a OFDM PHY layer, where each pull and push RE is made of 2 PRBs and 2 OFDM symbols.}
    \label{fig:frame}
\end{figure}

\glsreset{re}
\glsreset{prb}
\glsreset{phy}
The frame needs to be designed to accommodate both push- and pull-based communication modes. Accordingly, we employ an instance of the \gls{cfc} frame~\cite{pandey2025pullpushmag}, an example of which is given in Fig.~\ref{fig:frame}. Each frame has a fixed duration of $T$~seconds and follows an \gls{ofdm} structure. Overall, a total of $R$ \glspl{re} are available for \gls{ul} transmissions, and the frame is divided into pull and push subframes.
The pull subframe is divided in two: first, the \gls{bs} transmits a \gls{dl} \emph{pull scheduling signal}, composed of a \gls{ss}, the acknowledgments for messages in the previous frame, and the (multi-cast) requests for scheduling specific node transmissions. The scheduled nodes transmit orthogonally over the $Q$ pull \glspl{re}, being $Q \in \{0,\dots,R\}$~\cite{pandey2025pullpushmag}. The set of nodes scheduled in the $k$-th pull subframe is denoted as $\mc{S}(k) \subseteq \mc{N}_d$, being $|\mc{S}(k)| = Q$. Throughout the paper, we will present an analysis of the effect of the selection of $Q$ on performance and propose practical algorithms to manage it dynamically. 

Similarly, the push subframe begins with a \gls{dl} \emph{push control signal}, which contains another \gls{ss} and an \gls{aoii} threshold value $\theta^*(k)$: nodes that observed anomalies, but whose \gls{aoii} is lower than the threshold, will remain silent to avoid collisions with higher-priority traffic~\cite{kriouile2021minimizing}. Further details on the threshold mechanism are provided in Sec.~\ref{sec:scheduler-push}. Finally, the control signal includes the number  $P = R - Q$ of available push \glspl{re}. Accordingly, the devices that observe an anomaly at the beginning of the $k$-th frame, i.e., $\forall n\in\mc{N}_a: y_n(k) = 1$, listen for the control signal and decode it to recover slot boundaries and the frame structure.
The nodes whose \gls{aoii} status is compliant with the one requested by the \gls{bs}, i.e., $\Theta_n(k) \ge \theta^*(k)$, then transmit in a grant-free manner over the $P$ push \glspl{re}. Transmission follows a threshold-based \emph{\gls{fsa}} scheme~\cite{yue2023age}: every device whose \gls{aoii} is over the threshold randomly selects one of the $P$ \glspl{re} over which to transmit.

We consider nodes to be able to perform power control, so that communication takes place over a \emph{collision channel}. No collision occurs during the pull subframe due to the scheduled transmissions, and pull-based transmission is considered \emph{error-free} throughout the paper. Conversely, all collided packets over the same \gls{re} during the push subframe are erased.

The proposed structure is highly flexible: each \gls{re} can consist of an arbitrary number of \gls{ofdm} symbols and \glspl{prb} depending on the \gls{phy} layer performance required, making the proposed system easily compatible with \gls{3gpp} standards. More sophisticated contention resolution techniques, e.g., involving \gls{sic}, could be implemented without impacting the proposed system, and the model can be easily extended to a case with wireless channel erasure, as erased packets can be simply retransmitted in the next frame with limited consequences on \gls{dt} performance.

\subsection{Framework Architecture}

\begin{figure}[tb]
    \centering
    \includegraphics[width=0.87\columnwidth]{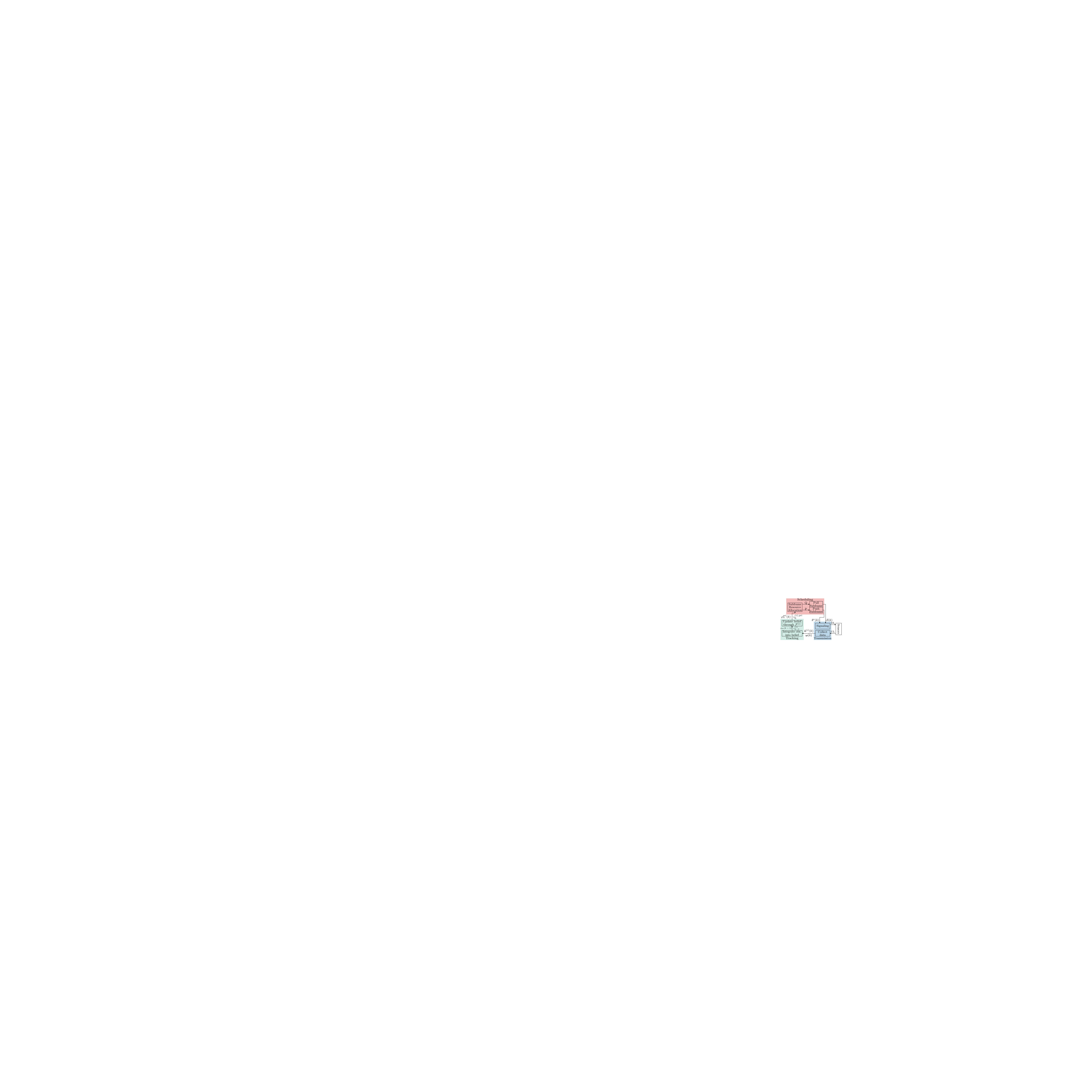}
    \caption{Block diagram of the proposed framework. Tracking, scheduling, and transmission modules acts to track and report \gls{dt} drift and anomalies.}
    \label{fig:diagram}
\end{figure}

The proposed framework consists of three key modules: \emph{transmission}, \emph{tracking}, and \emph{scheduling}, as shown in Fig.~\ref{fig:diagram}. For simplicity's sake, we assume it to run at the \gls{bs}, which interacts with the \gls{dt} to correct drifts and with an external agent to report anomalies. Alternative deployments (e.g., as an xApp) are possible but not investigated in this paper.

The transmission module handles \gls{dl} signaling and \gls{ul} packet reception, as described in Sec.~\ref{ssec:system_comms}. It also processes the collected data into observations compatible with the tracking module.

The tracking module monitors \gls{dt} drift events and anomalies by continuously updating a \gls{map} probability distribution for both phenomena. After each step from frame $k-1$ to frame $k$, it updates its belief by applying the transition function $\mc{F}^{(i)}$ introduced in Sec.~\ref{ssec:system_drift}. After the frame, the observation collected by the transmission module are integrated to refine the belief using the \gls{map} approach. Details on the belief computations are provided in Sec.~\ref{sec:tracking}. 

The scheduling module leverages the tracking module's belief distribution to reserve $Q$ \glspl{re} for pull and $P$ \glspl{re} for push subframes. It then uses the same belief to select the $Q$ sensors to pull from the set $\mc{N}_d$, i.e, to populate the set of scheduled nodes $S(k)$, and to set the anomaly \gls{aoii} threshold $\theta^*(k)$, which determines which sensors may attempt grant-free transmission. This threshold is designed to prioritize sensors with a high anomaly \gls{aoii} while managing collision risk. Further details are given in Sec.~\ref{sec:scheduler}.

\section{Network-Aware Digital Twin Update}
\label{sec:tracking}

\subsection{Digital Twin Alignment and Drift}
\label{sec:tracking-distributed}
As we described in Sec.~\ref{ssec:system_drift}, we consider the alignment of each \gls{dt} $i$ as an independent tracking model, over which the \gls{bs} maintains a belief distribution characterized by the \gls{pmf} $\zeta_k^{(i)}(x)$. We assume the initial state of the nodes belonging to cluster $i$ to be $x^{(i)}_0$, so the initial belief distribution is $\zeta_0^{(i)}(x)=\mc{I}\big(x=x^{(i)}_0 \big)$. After each transition from frame $k-1$ to frame $k$, we update the \emph{a priori} probability following transition probability function $\mc{F}^{(i)}$, which is known to the \gls{bs}:
\begin{equation}
   \zeta_k^{(i),\text{pri}}(x)=\sum_{x'\in\mc{X}^{(i)}}\zeta_{k-1}^{(i)}(x')\mc{F}^{(i)}\left(x';x\right). 
\end{equation}
Observations from sensor $n$ belong to set $\mc{O}_n$, while $\mc{O}^{(i)}=\prod_{n\in\mc{C}^{(i)}}\mc{O}_n$ for cluster $i$.
The \gls{bs} also knows the observation probability $\Omega:\mc{X}^{(i)}\times\mc{O}^{(i)}\to[0,1]$, which considers the sensors' imperfect measurements and potential errors.
The expected drift $\nu^{(i)}(k)$ is
\begin{equation}\label{eq:dist_risk}
    \nu^{(i)}(k)=\sum_{x\in\mc{X}^{(i)}}e^{(i)}(x)\zeta_k^{(i),\text{pri}}(x).
\end{equation}

However, sensor $n$ may fail to report during frame $k$, an outcome denoted as $\chi$, either because its packet collided (if it transmitted in the push subframe) or because it did not transmit anything; otherwise, it provides its observation $o_n\in\mc{O}_n$. The overall outcome for sensor cluster $i$ is $\tilde{\mb{o}}^{(i)}(k)\in\tilde{\mc{O}}^{(i)}=\prod_{n\in\mc{C}^{(i)}}\left(\mc{O}_n\cup\chi\right)$. 
We define the set $\mc{L}(\tilde{\mb{o}})$ of complete observations compatible with the received packets. This set represents the observation vectors that complete the outcome, i.e.:
\begin{equation}\label{eq:comp_obs}
    \mc{L}\Big(\tilde{\mb{o}}^{(i)}(k)\Big)=\left\{\mb{o}{\in}\mc{O}^{(i)}: o^{(i)}_n(k)\in\{o_n,\chi\}\,\forall n{\in}\mc{C}^{(i)}\right\}.
\end{equation}
Basically, set~\eqref{eq:comp_obs} contains all the possible observation vectors $\mb{o}\in\mc{O}^{(i)}$ compatible with the outcome of cluster $i$ at frame $k$. Reusing Example~\ref{example}, if node $1$ and $2$ of cluster $1$ provide observations interpreted as a drift by the \gls{bs} (labeled as $1$) but node $3$ and $4$ are silent on frame $k$, then $\mc{L}(\tilde{\mb{o}}^{(1)}(k))= \{[1,1,0,0]\T, [1,1,0,1]\T, [1,1,1,0]\T,[1,1,1,1]\T\}$.
We denote the probability of obtaining outcome $\tilde{\mb{o}}$ when the full observation was $\mb{o}$ as $p_{\chi}(\tilde{\mb{o}}|\mb{o})$. The probability of obtaining outcome $\tilde{\mb{o}}$ if the system is in state $x$ is $\tilde{\Omega}(x,\tilde{\mb{o}})=\sum_{\mb{o}\in\mc{L}(\tilde{\mb{o}})} p_{\chi}(\tilde{\mb{o}}|\mb{o})\Omega(x,\mb{o})$.

The tracking model does not use the existence of a transmission as a meaningful observation, as individual nodes do not know whether their \gls{dt} is aligned. We provide the update rule for the estimate when states are discrete:
\begin{equation}\label{eq:dist_post}    \zeta^{(i)}_k(x;\tilde{\mb{o}})=\frac{\zeta_k^{(i),\text{pri}}(x)\tilde{\Omega}(x,\tilde{\mb{o}})}{\sum_{x'\in\mc{X}^{(i)}}\zeta_k^{(i),\text{pri}}(x')\tilde{\Omega}(x,\tilde{\mb{o}})}.
\end{equation}
Observing values from more sensors in the cluster results in a higher degree of confidence. 

\subsubsection{HMM Scenario} In Scenario~\ref{scen:hmm}, observations are always either correct or missing, i.e.,
\begin{equation}
\tilde{\Omega}\left(x^{(i)}, \tilde{\mb{o}}^{(i)}\right)=\prod_{n\in\mc{C}^{(i)}}\left(\mc{I}(\tilde{o}_n=x_n)+\mc{I}(\tilde{o}_n=\chi)\right).
\end{equation}
The probability function $\tilde{\Omega}$ can be computed by cluster, as we assume no correlation between the observations of different clusters.
The expected drift can be computed as $\nu^{(i)}(k)=\sum_{x\in\mc{X}_d^{(i)}}\zeta_k^{(i),\text{pri}}(x)$,
as the risk function is equal to $1$ if the state belongs to $\mc{X}_d^{(i)}$ and $0$ otherwise. 

\subsubsection{Kalman Scenario} In Scenario~\ref{scen:kalman}, the observation \gls{pdf} $\Omega$ is Gaussian, with zero mean and a known covariance matrix. As the belief distribution is over a continuous variable or tensor $x$, the sums are simply replaced by integrals, and $\zeta_k^{(i)}(x)$ becomes a \gls{pdf}. Missing observations can be dealt with by the filter by setting specific rows of matrix $\mb{H}^{(i)}$ to $0$.
The expected drift can also be computed easily, as the Kalman filter implicitly keeps track of the drift. The expected \gls{mse} is equal to the trace of the prediction covariance matrix, i.e., $\nu^{(i)}(k)=\text{tr}\left(\mb{P}^{(i),\text{pri}}(k)\right)$.

\subsection{Anomaly Tracking} 
\label{sec:tracking-localized}

Consider a push-based threshold policy, in which nodes transmit if their \gls{aoii} is higher than a value $\theta^*(k)$. The \gls{bs} can maintain a belief distribution over the \gls{aoii} $\Theta_n(k)$ of each node $n$ after frame $k$, characterized by its \gls{pmf} $\varphi_n(\theta;k)$. Since the system starts with no anomalies, the \gls{bs} knows that the \gls{aoii} is equal to $0$ for all nodes, i.e., $\varphi_n(\theta;0)=\mc{I}(\theta=0)$, $\forall n\in\mc{N}$. If node $n$ successfully transmitted in frame $k$, the \gls{aoii} after that frame is equal to $0$ with probability $1$. On the other hand, if $n\notin\mc{S}(k)$, the prior belief $\varphi_n^{\text{pri}}(\theta;k)$ after frame $k$ is computed as:
\begin{equation} \label{eq:prior-belief-localized}
    \varphi_n^{\text{pri}}(\theta;k)=\begin{cases}
    (1-\lambda_n-\mu_n)\varphi_n(0;k-1)+\mu_n,&\theta=0,\;\\
    \lambda_{n}\varphi_n(0;k-1),&\theta=1;\\
    (1-\mu_n)\varphi_n(\theta-1;k-1),&\theta>1.
    \end{cases}
\end{equation}
If node $n$ has been scheduled during the pull subframe, it will report its measurement and its \gls{aoii} will be reset to $0$ before the beginning of the push subframe. In other cases, the prior belief of frame $k$ for $\theta = 1$ is the belief of the anomaly being generated in the past frame, which requires \gls{aoii} at the previous step to be $0$. On the other hand, we can have $\theta=0$ if there either was no anomaly in the previous frame, and a new anomaly was not generated during that interval, or if a previous anomaly spontaneously resolved itself. Finally, for $\theta > 1$, the \gls{pmf} of the prior belief in frame $k$ is the shifted version of the one in frame $k-1$, multiplied by $(1-\mu_n)$, i.e., the probability of the anomaly not resolving itself.

We consider vector $\bm{\omega}(k)\in\left(\mc{N}_a\cup\{-1,0\}\right)^{P}$, which
contains the outcome of all \glspl{re} in the $k$-th push subframe. The outcome of an \gls{re} is $0$ if no node transmitted and $-1$ if a collision happened, i.e., more than one node attempted to transmit over the same \gls{re}. If a node successfully transmitted in the \gls{re}, the outcome corresponds to its index $n$. 

Then, there are three cases. Case $1$) If node $n$ successfully transmitted, i.e., if $n\in\bm{\omega}(k)$, its \gls{aoii} is reset to $0$ with probability $1$, and we have $\varphi_n(\theta;k+1)=\mc{I}(\theta=0)$. Case $2$) Node $n$ did not attempt to transmit, and there were no collisions in the frame, i.e., $\{-1,n\}\cap\bm{\omega}(k)=\varnothing$, so the \gls{bs} knows that node $n$ was silent. As such, we know that $\Theta_n(k)\leq\theta^*(k)$, or node $n$ would have transmitted. The belief is updated as
\begin{equation}
\varphi_n(\theta;k)=\frac{\varphi_n^{\text{pri}}(\theta;k)\mc{I}\left(\theta\leq\theta^*(k)\right)}{\sum_{j=0}^{\theta^*(k)}\varphi_n^{\text{pri}}(j;k)},\ \{-1,n\}\cap\bm{\omega}(k)\!=\!\varnothing,
\end{equation}
as values higher than $\theta^*(k)$ are not compatible with the outcome of the subframe. The prior belief is normalized for the  feasible values of $\theta$, as we have no information aside from the fact that node $n$'s \gls{aoii} is lower than or equal to the threshold $\theta^*(k)$. Case $3$) Whenever there is at least one collision, things are more complicated, as the \gls{bs} cannot know which nodes transmitted or even how many, as more than $2$ nodes may have collided over the same \gls{re}. Under threshold $\theta^*(k)$, the estimated transmission probability for node $n$ is
\begin{equation}
\alpha_n\left(\theta^*(k);\varphi_n^{\text{pri}}(k)\right)=\sum_{j=\theta^*(k)+1}^{\infty}\varphi_n^{\text{pri}}(j;k).
\end{equation}
We also define the set of potentially active nodes $\mc{A}(\theta^*(k))$, whose cardinality is $A(\theta^*(k))$:
\begin{equation}
    A(\theta^*(k))=\sum_{n\in\mc{N}_a} \mc{I}\left(\alpha_n\left(\theta^*(k);\varphi_n^{\text{pri}}(k)\right)>0\right).
\end{equation}
\begin{approximation}\label{appr:prob}
    In the following, we assume that all nodes have the same activation probability
    \begin{equation}
        \bar{\alpha}(k)=\frac{\sum_{n\in\mc{N}_a}\alpha_n\left(\theta^*(k);\varphi_n^{\text{\emph{pri}}}(k)\right)}{A(\theta^*(k))}.
    \end{equation}
\end{approximation}
This approximation greatly simplifies our calculations, and is usually close to reality in practice, as the belief update will lead to very similar outcomes.
The probability of there being $a$ active nodes in a push subframe is
\begin{equation}
    \Pr(a|\theta^*(k))\simeq\binom{A(\theta^*(k))}{a}\left(1-\bar{\alpha}(k)\right)^{A(\theta^*(k))-a}\left(\bar{\alpha}(k)\right)^a.
\end{equation}

The outcome of the push subframe includes $s$ successful transmissions, $c$ collisions, and $P-s-c$ empty \glspl{re}.\footnote{In the following, we will omit the condition on the value of $P$ from all probability calculations for the sake of readability.} As mentioned above, we have no information on the number and the indexes of the nodes that have transmitted in the $c$ \glspl{re} experiencing a collision.

\begin{approximation}\label{appr:threeway}
    No collision involves more than $3$ nodes.
\end{approximation}
This second approximation allows us to avoid considering very rare cases in which several nodes collide over the same \gls{re}, whose probability is almost negligible, while still maintaining a non-trivial belief. We compute the likelihood of the outcome, conditioning on the total number of active nodes:
\begin{equation}
\begin{aligned}
    \Pr\left(s,c|a\right)\simeq&\binom{a}{s,c,c}\frac{P!}{(P-s-c)!c!P^a2^c3^{a-s-2c}}.
\end{aligned}
\end{equation}
where $\binom{n}{m_1,\ldots,m_k}$ is the multinomial coefficient~\cite[p. 77]{aigner2012combinatorial}.
The probability is computed by selecting $s$ successful nodes, then $c$ \glspl{re} and placing $2$ nodes over each one. The remaining $a - s - 2c$ nodes can simply be placed on any of the $c$ \glspl{re} in which there is already a collision. We can apply Bayes' theorem to compute the \emph{a posteriori} probability of $a$ nodes being active after observing $s$ successes and $c$ collisions in the push subframe:
\begin{equation}
    \Pr\left(a|s,c;\theta^*(k)\right)\simeq\frac{\Pr(a|\theta^*(k))\Pr\left(s,c|a\right)}{\sum_{a=s+2c}^A \Pr\left(s,c|a\right)},\ a\geq s+2c.
\end{equation}
Naturally, the probability for $a<s+2c$ is $0$, as the outcome is impossible for that number of active nodes. We can compute the probability of event $\chi_n$, which indicates that node $n$ was among the colliding nodes:
\begin{equation}
    \Pr\left(\chi_n|s,c;\theta^*(k)\right)\simeq\sum_{a=s+2c}^{A(\theta^*(k))}\frac{(a-s)\Pr\left(a|s,c;\theta^*(k)\right)}{A(\theta^*(k))-s},
\end{equation}
which is the same for each $n\in\mc{A}(\theta^*(k))$ due to Approx.~\ref{appr:prob}.

\begin{approximation}
    We consider belief updates over individual nodes, rather than over the joint probability distribution.
\end{approximation}
This further approximation is necessary due to the combinatorial nature of the problem, as the network may include tens or hundreds of nodes. As we are generally dealing with rare events (only a few nodes are colliders, while most will not transmit), the error caused by the approximation is also usually negligible in practice.
We provide the \emph{a posteriori} belief over the \gls{aoii} of each node by applying Bayes' theorem:
\begin{equation}
    \varphi_n(\theta;k+1){=}\!\begin{cases}
    \displaystyle \frac{\varphi_n^{\text{pri}}(\theta;k)\big(1{-}\Pr\!\big(\chi_n|s,c;\theta^*(k)\big)\big)}{\sum_{j=0}^{\theta^*(k)}\varphi_n^{\text{pri}}(j;k)},\!&\!\theta{\leq}\theta^*(k);\\
    \displaystyle \frac{\varphi_n^{\text{pri}}(\theta;k)\Pr\!\big(\chi_n|c,s;\theta^*(k)\big)}{\sum_{j=\theta^*(k)+1}^{\infty}\varphi_n^{\text{pri}}(j;k)},\! &\!\theta{>}\theta^*(k).
    \end{cases}
\end{equation}
We can thus update all nodes that did not transmit successfully in a push subframe with at least one collision.\footnote{While the approximations we introduced are used in the belief update, all simulations were performed without these assumptions.}


\section{Joint Push-Pull Scheduling}
\label{sec:scheduler}

\subsection{Pull Subframe}
\label{sec:scheduler-pull}

In the pull subframe, we follow a simple principle: the \gls{bs} should maximize the reduction of the drift that can be gained from the available \glspl{re}. We adopt an iterative approach: we start from an empty schedule $\mc{S}_0=\emptyset$. We define set  $\mc{S}_{\ell}^{(i)}$ for each cluster $\mc{C}^{(i)}$, which contains the nodes that have been scheduled for cluster $i$ after iteration $\ell$; naturally, $\mc{S}^{(i)}_0=\emptyset\ \forall i$. Unscheduled nodes will not report during frame $k$, unless they transmit in the push subframe; thus, the set of possible pull outcomes $\mc{M}\left(\mc{S}_{\ell}^{(i)}\right)$ for scheduling set $\mc{S}_{\ell}^{(i)}$ is\footnote{Following Example~\ref{example}, if $\mc{S}^{(1)}_\ell = \{1,2\}$ then $\mc{M}(\mc{S}^{(1)}_\ell)= \{[0,0,\chi,\chi]\T, [0,1,\chi,\chi]\T, [1,0,\chi,\chi]\T,[1,1,\chi,\chi]\T\}$.}
\begin{equation}
\mc{M}\Big(\mc{S}_{\ell}^{(i)}\Big)=\Big\{\tilde{\mb{o}}\in\tilde{\mc{O}}^{(i)}: \tilde{o}_n=\chi\,\forall n\notin\mc{S}_{\ell}^{(i)}\Big\}.
\end{equation}
We define the \emph{a posteriori} expected drift function after receiving observation $\tilde{\mb{o}}$ as
\begin{equation}\label{eq:exp_drift}    \hat{e}_k\left(\mc{S}_{\ell}^{(i)}\right)=\sum_{{\tilde{\mb{o}}\in\mc{M}\left(\mc{S}_{\ell}^{(i)}\right)}} \frac{ \sum_{{x\in\mc{X}^{(i)}}}\zeta_k^{(i),\text{pri}}(x)\tilde{\Omega}(x,\tilde{\mb{o}})e^{(i)}(x)}{\sum_{{x'\in\mc{X}^{(i)}}}\zeta_k^{(i),\text{pri}}(x')\tilde{\Omega}(x',\tilde{\mb{o}})}.
\end{equation}
This equation simply corresponds to the expected value of the drift function for each possible observation, weighted by the prior probability of obtaining that observation.
We perform a look-ahead step, considering the gain when adding a generic node $n$, belonging to cluster $i$, to the schedule:
\begin{equation}\label{eq:info_gain}
\begin{aligned}         G\left(n;\mc{S}_{\ell}\right)=\hat{e}_k\left(\mc{S}_{\ell}^{(i)}\cup\{n\}\right)-\hat{e}_k\left(\mc{S}_{\ell}^{(i)}\right).
\end{aligned}
\end{equation}
We can then sort all unscheduled nodes by their information gain and schedule the node with the highest information gain:
\begin{equation}
    \mc{S}_{\ell+1}=\mc{S}_{\ell}\cup\left\{\argmax_{n\in\mc{N}_d\setminus\mc{S}_{\ell}} G(n;\mc{S}_{\ell})\right\}.
\end{equation}
Considering the fact that node $n^*$ has been scheduled, we recompute~\eqref{eq:info_gain} for its cluster. We iterate this operation until all $Q$ \glspl{re} in the pull subframe are assigned to a node, obtaining the final schedule $\mc{S}(k)=\mc{S}_Q$.

\subsubsection{HMM} In Scenario~\ref{scen:hmm}, we use the entropy of $\hat{e}_k\left(\mc{S}^{(i)}\right)$ to compute the information gain, as minimizing the entropy corresponds to minimizing the uncertainty over the drift state. We can apply~\eqref{eq:exp_drift} directly, and as $e^{(i)}(x)$ is binary, we get
\begin{equation}
\hat{e}_k\left(\mc{S}^{(i)}\right)=\sum_{{\tilde{\mb{o}}\in\mc{M}\left(\mc{S}^{(i)}\right)}} \frac{\sum_{{x\in\mc{X}_d^{(i)}}}\zeta_k^{(i),\text{pri}}(x)\tilde{\Omega}(x,\tilde{\mb{o}})}{\sum_{{x'\in\mc{X}^{(i)}}}\zeta_k^{(i),\text{pri}}(x')\tilde{\Omega}(x',\tilde{\mb{o}})}.
\end{equation}
Furthermore, we know that $\tilde{\Omega}(x',\tilde{\mb{o}})$ is equal to $1$ if all observations that are not missing are identical to the corresponding elements of the state, and $0$ otherwise.

\subsubsection{Kalman} In Scenario~\ref{scen:kalman}, the Kalman filter update equation allows us to compute the trace of the covariance matrix if we know $\mc{S}^{(i)}$, so we can use it in~\eqref{eq:info_gain} without solving~\eqref{eq:exp_drift} explicitly. 
The steps for the information gain calculation  are equivalent to the \gls{mse} scheduler of~\cite[Sec.~III-A]{chiariotti2022scheduling}.

\subsection{Push Subframe}
\label{sec:scheduler-push}
The objective for the push subframe is to allow nodes with a high \gls{aoii} (relative to the rest of the network) to transmit as quickly as possible, while avoiding collisions. Collisions are more harmful than in standard \gls{fsa}, as they have the additional effect of reducing the information that the \gls{bs} can use to improve its estimate of nodes' \gls{aoii}. Therefore, we set a maximum collision probability $\sigma$.
We define the highest possible \gls{aoii} $\theta_{\max}(k)$ across the whole network as
\begin{equation}
\theta_{\max}(k)=\max_{n\in\mc{N}_a}\sup\left\{\theta:\varphi_n^{\text{pri}}(\theta;k)>0\right\}.
\end{equation}
We then compute the probability $\Pr\left(\chi|\theta^*\right)$ of a collision occurring with threshold $\theta^*$. Exploiting Approx.~\ref{appr:prob}, we get
\begin{equation}
  \Pr\left(\chi|\theta^*\right)=\sum_{a=2}^{A(\theta^*)}\left(1-\frac{P!}{P^a(P-a)!}\right)\Pr\left(a|\theta^*\right).
\end{equation}
We define set $\mc{V}(k)$ as the set of possible thresholds that have a collision probability lower than a threshold value $\sigma$:
\begin{equation}
    \mc{V}(k)=\left\{\theta^*\in\{0,\ldots,\theta_{\max}(k)-2\}:\Pr\left(\chi|\theta^*\right)\leq\sigma\right\}.
\end{equation}
The transmission threshold is selected as
\begin{equation}
    \theta^*(k)=\begin{cases}
        \sup(\mc{V}(k)), &\text{if }\mc{V}(k)\neq\varnothing;\\
        \max(0,\theta_{\max}(k)-2), &\text{otherwise.}
    \end{cases}
\end{equation}

\subsection{Subframe Resource Allocation}
\label{sec:scheduler-management}

The balance between the pull-based and push subframes is a critical parameter in the system: the two subframes need to share a finite pool of resources, and striking a balance between our two objectives is a fundamental problem.

We can provide a natural measurement of the urgency of pull transmissions, given by the expected drift $\nu^{(i)}(k)$ obtained with~\eqref{eq:dist_risk}. We can define a maximum drift $\nu_{\max}$ and determine the drift urgency as the drift relative to $\nu_{\max}$:
\begin{equation}
    \bar{\nu}(k)=\min\bigg(1,D^{-1}\sum_{i=1}^D\frac{\nu^{(i)}(k)}{\nu_{\max}}\bigg).
\end{equation}

We need to define an urgency metric for anomalies. We consider a maximum acceptable \gls{aoii} $\tilde{\theta}$ and define the push urgency as the average \gls{aoii} violation probability:
\begin{equation}
\varepsilon(k)=(|\mc{N}_a|)^{-1}\sum_{n\in\mc{N}_a}\sum_{\theta=\tilde{\theta}+1}^{\infty}\varphi^{\text{pri}}_n(\theta;k).
\end{equation}

The balance between \gls{dt} alignment and anomaly detection can be considered using $\bar{\nu}(k)$ and $\varepsilon(k)$ as proxies. We propose two dynamic allocation schemes, called \emph{\gls{rsm}} and \emph{\gls{ssm}}. The former aims at quickly adapting the resource allocation to the current urgency, while the latter changes resource allocation slowly over multiple frames. In both cases, we set a minimum amount of resources $R_{\min}$ that are always allocated to each subframe to avoid resource starvation.

The \gls{rsm} resource allocation scheme allocates \glspl{re} for the push subframe following this simple rule:
\begin{equation}
    P_{\text{RSM}}(k)=\left\lfloor\frac{R\varepsilon(k)}{\varepsilon(k)+\bar{\nu}(k)}\right\rfloor.
\end{equation}
Then, $Q_{\text{RSM}} (k) = R - P_{\text{RSM}}(k)$.
On the other hand, the \gls{ssm} scheme only changes the allocation from the previous subframe by $1$ \gls{re}:
\begin{equation}
    P_{\text{SSM}}(k)=\begin{cases}
        P_{\text{SSM}}(k-1)+1, &\text{if }\varepsilon(k)-\bar{\nu}(k)>\eta;\\        
        P_{\text{SSM}}(k-1)-1, &\text{if }\bar{\nu}(k)-\varepsilon(k)>\eta;\\
        P_{\text{SSM}}(k-1), &\text{otherwise.}
    \end{cases}
\end{equation}
where $\eta$ is a hysteresis parameter to avoid too many fluctuations. As before, $Q_{\text{SSM}}(k) = R - P_{\text{SSM}}(k)$. In both cases, results are restricted to interval $\{R_{\min},R_{\min}+1,\ldots,R-R_{\min}\}$.

\section{Numerical Evaluation}\label{sec:results}

\addtolength{\tabcolsep}{-0.5em}
\begin{table}[b]
    \centering
    \caption{Main parameters.}
    \scriptsize
    \begin{tabular}{|ccc|ccc|}
    \hline
    \textbf{Param.} & \textbf{Meaning} & \textbf{Value} & \textbf{Param.} & \textbf{Meaning} & \textbf{Value}\\
    \hline
    \multicolumn{3}{|c|}{\textbf{Frame}} & \multicolumn{3}{c|}{\textbf{Nodes}} \\
    \hline
    $R$ & \glspl{re} per frame &$20$ & $N_d$ & \gls{dt} sensor nodes & $80$\\
    $Q$ & \glspl{re} for pull subframe & 10 & $N_a$ & Anomaly sensor nodes & $100$ \\ 
    $P$ & \glspl{re} for push subframe & 10 & $\mc{N}_a\cap\mc{N}_d$& Sensors & $\mc{N}_d$ \\
    $T$ & Frame duration & $10$~ms & $C$ & Nodes per \gls{dt} cluster & $4$\\
    \hline
    \multicolumn{3}{|c|}{\textbf{Scheduling}} &  \multicolumn{3}{c|}{\textbf{Simulations}} \\
    \hline
    $\sigma$ & \gls{pps} collision thr. & $0.2$ & $N_{\text{ep}}$ & Monte Carlo episodes& $100$ \\
    $\nu_{\text{max}}$ & Drift urgency threshold & $5$ & 
    $K_{\text{ep}}$ & Episode duration& $10$~s\\
    $\tilde{\theta}$ & Anomaly \gls{aoii} risk thr. & $20$~ms &  $G_{\text{FSA}}$ & FSA max load & $0.9$ \\
    $R_{\min}$ & Min. subframe \glspl{re} & $5$ &
    $G_{\min(\max)}$ & AFSA min/max load & $0.2$/$0.95$ \\
    $\eta$ & \gls{ssm} hysteresis thr. & $0.005$ &
    $\gamma$ & AFSA discount rate & $0.1$ \\
    \hline
    \multicolumn{6}{|c|}{\textbf{Scenarios}} \\    
    \hline
    $\rho_a$ & Anomaly rate & \multicolumn{1}{c}{$3$~Hz} & $\mu_n$ & Anomaly resolution rate & $0$\\
    $\nu_\text{reset}$ & \gls{dt} reset thr. (\gls{hmm}) & \multicolumn{1}{c}{$0.95$} & 
    $\rho_d$ & \gls{dt} drift rate (\gls{hmm}) & $3$~Hz \\
    $\sigma_v$ & Obs. SD (Kalman) & \multicolumn{1}{c}{$0.25$} & $\sigma_w$ & Process SD (Kalman) & $0.5$ \\
    \hline
    \end{tabular}    
    \label{tab:param}
\end{table}


In order to evaluate \gls{pps}, we performed Monte Carlo simulations over $N_{\text{ep}}=100$ episodes of $K_{\text{ep}}=1000$ frames each. The duration of an episode is enough to observe steady-state behavior, and the number of episodes was chosen to avoid distorting effect from individual runs.\footnote{The full simulation code used to generate the results in this work is available at \url{https://github.com/signetlabdei/push-pull-anomaly-tracking}.}
For the sake of simplicity, we consider clusters of a fixed size, i.e., $|\mc{C}^{(i)}|=C\, \forall i$. In the anomaly detection task, we consider anomalies not to resolve spontaneously, i.e., $\mu_n=0\,\forall n\in\mc{N}_a$. The frequency of anomalies at each node is upper bounded by the anomaly rate $\rho_a=(|\mc{N}_a|T)^{-1}\sum_{n\in\mc{N}_a}\lambda_n$. Also, we consider $\mc{N}_d\subset\mc{N}_a$, i.e., there are more nodes reporting anomalies than those involved in \gls{dt}, but all \gls{dt} sensors may also report anomalies. 

In the following, we first report the results for each subframe separately, evaluating the two components of \gls{pps} individually, then consider the full scheme and subframe resource allocation schemes.
The standard parameters used are reported in Table~\ref{tab:param}.

\subsubsection{HMM Scenario} To assess the performance of \gls{pps} in Scenario~\ref{scen:hmm}, we consider a simple Markov chain in which each sensor has a binary state and transitions independently, so the drift state of the system is the row vector $\mb{x}^{(i)}(k)\in\mc{X}^{(i)}=\{0,1\}^{C}$. We consider the \gls{dt} to be drifting if a majority of nodes in the cluster is in state $1$, i.e., $
    \mc{X}_d^{(i)}=\left\{\mb{x}\in\{0,1\}^C:\sum_{n\in\mc{C}^{(i)}}x_n\geq\frac{C}{2}\right\}$.
Furthermore, once the system is in a drift state, we assume that nodes in state $1$ will remain in that state until the drift is reported. The transition probability is defined by matrix $\mb{U}^{(i)}$, which can be computed from the individual nodes' transition probabilities.
Additionally, since measurements are error-free and binary, the set of states that are compatible with observation $\mb{o}^{(i)}(k)$, defined in~\eqref{eq:comp_obs}, becomes $
    \mc{L}\left(\mb{o}^{(i)}(k)\right)=\left\{\mb{x}\in\{0,1\}^C:o_n(k)\in\{x_n,\chi\}\,\forall n\in\mc{C}^{(i)}\right\}$.
 
In the following, we use the \gls{aoii} as our main performance metric, as it measures the time since the \gls{dt} started drifting:
\begin{equation}
    \Psi^{(i)}(k)=k-\sup\left\{\ell\in\{0,\ldots,k\}:z^{(i)}(\ell)=0\right\}.
\end{equation}
Thus, we report the \gls{aoii} as a function of the drift rate $\rho_d=\Big(T\E{\inf\Big(t:z^{(i)}(k)=1|x^{(i)}(0)=x_0^{(i)}\Big)}\Big)^{-1}$, which corresponds to the average frequency of drift events that each cluster generates.
As drift is solved when $\nu^{(i)}(k)>\nu_{\text{reset}}$, this is the maximum possible rate of drift events, i.e., the expected absorption rate of the cluster Markov chain when starting from $x_0^{(i)}$, considering all drift states as absorbing. 

\subsubsection{Kalman Scenario} To evaluate the performance of \gls{pps} in Scenario~\ref{scen:kalman}, we consider a system in which $M=C$, i.e., the dimension of the state corresponds to the number of sensors, and $\mb{H}=\mb{I}_C$ for all clusters. We define the system update matrix $\mb{F}$ as follows:
 \begin{equation} \label{eq:F-well}
     F_{i,j}=\begin{cases}
         1, &\text{if }i=j;\\
         -\frac{1}{9}, &\text{if }|\text{mod}(i-j, C)|=1;\\
         0, &\text{otherwise,}
     \end{cases}
 \end{equation}
 where $\text{mod}(m,n)$ is the integer modulo operation. Then, we assume $\mb{\Sigma}_w=\sigma_w^2 \, \mb{I}_C$ and $\mb{\Sigma}_v=\sigma_v^2 \, \mb{I}_{C}$. In the following, we keep the observation \gls{sd} $\sigma_v=0.25$ and vary the process \gls{sd} $\sigma_w\in[0,1]$ to analyze different systems.
We also consider a mis-specified system, using an estimated update matrix $\hat{\mb{F}}$ in the \gls{dt} model:
  \begin{equation} \label{eq:F-mis}
     \hat{F}_{i,j}=\mc{I}(i=j)-\left(\frac{1}{9}-\beta_F\right)\mc{I}\left(|\text{mod}(i-j, C)|=1\right).
  \end{equation}
 with $\beta_F\in\left[-\frac{1}{9},\frac{1}{9}\right]$. We note that the right limit of the interval turns $\hat{\mb{F}}$ into the identity matrix, and thus strongly affects the dynamic system. In the following, we use the \gls{mse} as our main performance metric, as it measures the error between the ground truth and estimated drift state: $e(\mb{x}, \hat{\mb{x}})$.

\subsection{Benchmark Algorithms}

We consider two benchmark algorithms for \gls{dt} drift: first, we use the common \gls{maf}~\cite{bedewy2017age} benchmark, which selects the nodes with the highest \gls{aoi} in each frame. Conversely, the \gls{cra} scheme attempts to minimize the expected drift, exploiting $\nu^{(i)}(k)$. 
To ensure that drift is detected quickly, \gls{cra} sorts clusters from the highest $\nu^{(i)}(k)$ to the lowest. All nodes in the highest-drift unscheduled cluster are scheduled if at least $C$ of the $Q$ \glspl{re} in the pull subframe are still free. When the number of remaining \glspl{re} is less than $C$, they are filled by randomly selecting nodes from the next cluster.

In addition to \gls{maf}, we also consider two state-of-the-art random access algorithms for push subframes, pure \gls{fsa} and \gls{afsa}. \gls{maf} is often used as a benchmark for random access policies, and can perform well 
under heavy loads, as it can schedule packets orthogonally. In pure \gls{fsa}~\cite{yue2023age}, nodes with an anomaly transmit with probability $p_{\text{tx}}$ until it is resolved. We tune $p_{\text{tx}}$ to obtain a target load $G_{\text{FSA}}$ under anomaly rate $\rho_a$:
\begin{equation}
    p_{\text{tx}}(k)=(N_a\rho_a)^{-1} T P G_{\text{FSA}}.
\end{equation}
Instead, \gls{afsa} is a protocol inspired by the adaptive \gls{fsa} literature~\cite{khandelwal2007asap,wang2024age} that follows the rate adaptation procedure: 
\begin{equation}
    p_{\text{tx}}(k)=p_{\text{tx}}(k-1)+P^{-1}\gamma(P_{\chi}-P_{\text{sil}}),
\end{equation}
where $\gamma$ is a discount parameter controlling the variability of the rate, and $P_{\chi}$ and $P_{\text{sil}}$ represent the number of \glspl{re} in the past frame that resulted in a collision and no transmission, respectively. \gls{afsa} adapts the transmission probability dynamically, backing off when there are multiple collisions and increasing the rate when most \glspl{re} are unused. 
Finally, to avoid resource starvation, we set a minimum load $G_{\min}$ and a maximum load $G_{\max}$, limiting the adaptation range of \gls{afsa}.

\subsection{Performance: DT Drift in the HMM scenario}


\begin{figure}
\centering
\subfloat[Average drift \gls{aoii} $\E{\Psi}$. \label{fig:pull_frame_hom_avg}]{\begin{tikzpicture}
\pgfplotstableread{fig/fig_data/pull_frame_hom_avg.csv}\homtable;
\pgfplotstableread{fig/fig_data/pull_frame_het_avg.csv}\hettable;
\begin{axis}[
    width=\fwidth,
    height=\fheight,
    ybar=0.5pt,
    tick align=inside,
    bar width=2pt,
    legend style={font=\scriptsize, at={(1, 1)}, anchor=north east, legend columns=3},
    xmin=4.5,
    xmax=15.5,
    xlabel={$Q$ (REs)},
    ymin=0,
    ymax=1,
    xtick={5,6,7,8,9,10,11,12,13,14,15},
    ytick={0,0.25,0.5,0.75,1,1.25, 1.5,1.75, 2},
    yticklabels={0,2.5, 5, 7.5, 10, 12.5, 15, 17.5, 20},
    ylabel={$\E{\Psi}$ (ms)},
    legend image code/.code={\draw [#1] (0cm,-0.1cm) rectangle (0.15cm,4pt); },
]

    \addplot[color0,fill={white!20!color0}] table[x=Q, y=MAF] {\homtable}; 
    \addlegendentry{MAF (hom.)};    
    \addplot[style={color2,fill={white!20!color2}}] table[x=Q, y=CRA] {\homtable}; 
    \addlegendentry{CRA (hom.)};
    \addplot[style={color4,fill={white!20!color4}}] table[x=Q, y=PPS] {\homtable}; 
    \addlegendentry{PPS (hom.)};
    \addplot[color0,fill={white!20!color0},
    postaction={pattern=north east lines}] table[x=Q, y=MAF] {\hettable}; 
    \addlegendentry{MAF (het.)};
    \addplot[color2,fill={white!20!color2},
    postaction={pattern=north east lines}] table[x=Q, y=CRA] {\hettable}; 
    \addlegendentry{CRA (het.)};
    \addplot[color4,fill={white!20!color4},
    postaction={pattern=north east lines}] table[x=Q, y=PPS] {\hettable}; 
    \addlegendentry{PPS (het.)};

\end{axis}
\end{tikzpicture}}\\ \vspace{-0.2cm}
\subfloat[$99$th percentile drift \gls{aoii} $\Psi_{99}$. \label{fig:pull_frame_hom_99}]{\begin{tikzpicture}
\pgfplotstableread{fig/fig_data/pull_frame_hom_99.csv}\homtable;
\pgfplotstableread{fig/fig_data/pull_frame_het_99.csv}\hettable;
\begin{axis}[
    width=\fwidth,
    height=\fheight,
    ybar=0.5pt,
    tick align=inside,
    bar width=2pt,
    xmin=4.5,
    xmax=15.5,
    xlabel={$Q$ (REs)},
    ymin=0,
    xtick={5,6,7,8,9,10,11,12,13,14,15},   
    ymax=10,
    ytick={0,2.5,5,7.5,10,12.5, 15, 17.5, 20},
    yticklabels={0,25,50,75,100,125,150,175,200},
    ylabel={$\Psi_{99}$ (ms)},
]

    \addplot[color0,fill={white!20!color0}] table[x=Q, y=MAF] {\homtable}; 
    \addplot[style={color2,fill={white!20!color2}}] table[x=Q, y=CRA] {\homtable}; 
    \addplot[style={color4,fill={white!20!color4}}] table[x=Q, y=PPS] {\homtable}; 
    \addplot[color0,fill={white!20!color0},
    postaction={pattern=north east lines}] table[x=Q, y=MAF] {\hettable}; 
    \addplot[color2,fill={white!20!color2},
    postaction={pattern=north east lines}] table[x=Q, y=CRA] {\hettable}; 
    \addplot[color4,fill={white!20!color4},
    postaction={pattern=north east lines}] table[x=Q, y=PPS] {\hettable}; 

\end{axis}
\end{tikzpicture}}
\caption{Average and worst-case \gls{aoii} for \gls{dt} drift as a function of $Q$ under a purely pull-based system with drift rate $\rho_d = 3$~Hz, considering homogeneous and heterogeneous clusters.}\vspace{-0.3cm}
\label{fig:pull_frame}
\end{figure}

We first focus on assessing \gls{dt} alignment performance in Scenario~\ref{scen:hmm}. In this scenario, performance depends on the overall drift rate $\rho_d$, but also on the state change probability of individual nodes in each cluster. Thus, we test the performance in a homogeneous and a heterogeneous cases. In the former, all nodes have the same probability $u_n(0,1)$ of moving from state $0$ to state $1$, while in the latter, the probabilities for the sensors in a cluster are scaled according to vector $(1, 7,7.25, 7.5)$. We have $u_4(0,1)=7.5u_1(0,1)$, i.e., the fourth sensor in the cluster flips to state $1$ $7.5$ times as often as the first. The probability of a sensor remaining in state $1$ if the \gls{dt} is still aligned is $u_n(1,1)=0.9$ for all clusters. 
Transition matrix $\mb{U}^{(i)}$ can be easily computed from these values~\cite{pinsky2010stochastic}. We normalize the vector $\mb{u}(0,1)$ to match the target rate $\rho_d$.

Fig.~\ref{fig:pull_frame} shows the performance of \gls{pps} in a purely pull-based system, i.e., $Q = R$, setting $\rho_d = 3$~Hz. Let us start analyzing homogeneous clusters:  as shown in Fig.~\ref{fig:pull_frame_hom_avg}, \gls{pps} outperforms \gls{maf} in terms of the average drift \gls{aoii}, for smaller values of $Q$--aside from the extreme with $Q=5$--while the gain is negligible for the other extreme, $Q=15$.
Conversely, \gls{cra} tends to underperform for lower values of $Q$, but works well for $Q\geq10$.
Finally, \gls{maf} and \gls{pps} have a similar performance if we consider the $99$th percentile \gls{aoii}, shown in Fig.~\ref{fig:pull_frame_hom_99}, while \gls{cra} is slightly worse. 
\gls{pps} has a larger advantage in the heterogeneous cluster scenario: its average \gls{aoii} in Fig.~\ref{fig:pull_frame_hom_avg} is almost $40\%$ lower than the one obtained by \gls{maf} and $50\%$ lower than the one obtained by \gls{cra} if $Q$ is small. While the gain diminishes for larger values of $Q$, remaining over $20\%$ for $Q=15$. In this case, \gls{cra} performs as well or better than \gls{maf} for $Q\geq10$. This is because \gls{maf} does not consider the relative risk of clusters and nodes, and is thus less useful for heterogeneous nodes. This general trend is confirmed for worst-case performance in Fig.~\ref{fig:pull_frame_hom_99}, in which \gls{pps} can often reduce the $99$th percentile \gls{aoii} by $10$~ms (i.e., one frame) with respect to \gls{maf} and \gls{cra}.

\begin{figure}
\centering
\subfloat[Average drift \gls{aoii} $\E{\Psi}$. \label{fig:pull_load_avg}]{\begin{tikzpicture}
\pgfplotstableread{fig/fig_data/pull_load_avg.csv}\loadedtable;
\begin{axis}[
    width=.42\fwidth,
    height=\fheight,
    xlabel={$\rho_d$ (Hz)},
    ylabel={$\E{\Psi}$ (ms)},
    xmin=1, xmax=5,
    ymin=0, ymax=0.6,    
    legend style={font=\scriptsize, at={(1,0.5)}, legend columns=1, anchor=west, draw=none, fill=none},
    ytick={0, 0.2, 0.4, 0.6},
    yticklabels={0,2,4,6}
]

\addplot[
    color=color0, mark=triangle,mark repeat=5
    ]
    table[x=abs_rate,y=MAF] {\loadedtable};
\addlegendentry{MAF}; 
\addplot[
    color=color2, mark=diamond,mark repeat=5, mark phase=3
    ]
    table[x=abs_rate,y=CRA] {\loadedtable};
\addlegendentry{CRA};
\addplot[
    color=color4, mark=square, mark repeat=5
    ]
    table[x=abs_rate,y=PPS] {\loadedtable};
\addlegendentry{PPS};

\end{axis}
\end{tikzpicture}}\hfill
\subfloat[$99$th percentile drift \gls{aoii} $\Psi_{99}$. \label{fig:pull_load_99}]{\begin{tikzpicture}
\pgfplotstableread{fig/fig_data/pull_load_99.csv}\loadedtable;
\begin{axis}[
    width=.5\fwidth,
    height=\fheight,
    xlabel={$\rho_d$ (Hz)},
    ylabel={$\Psi_{99}$ (ms)},
    xmin=1, xmax=5,
    ymin=0, ymax=10,
    legend pos=south east,
    ytick={0,2.5,5,7.5,10},
    yticklabels={0,25,50,75,100},
    legend style={legend columns=1, anchor=north west, at={(0.01,0.98)}}
]

\addplot[
    color=color0, mark=triangle,mark repeat=5
    ]
    table[x=abs_rate,y=MAF] {\loadedtable};
\addplot[
    color=color2, mark=diamond,mark repeat=5
    ]
    table[x=abs_rate,y=CRA] {\loadedtable};
\addplot[
    color=color4, mark=square, mark repeat=5
    ]
    table[x=abs_rate,y=PPS] {\loadedtable};

\end{axis}
\end{tikzpicture}}
\caption{Average and worst-case \gls{aoii} for \gls{dt} drift as a function of the drift rate $\rho_d$ under a purely pull-based system with $Q = 10$.}\vspace{-0.3cm}
\label{fig:pull_load}
\end{figure}

Then, we consider performance in the heterogeneous scenario as a function of the drift rate $\rho_d\in[1,5]$~Hz, setting a fixed number of \glspl{re} $Q=10$: Fig.~\ref{fig:pull_load} shows the average and worst-case \gls{aoii} for the three schemes. The gain in terms of expected \gls{aoii} increases with the load, as Fig.~\ref{fig:pull_load_avg} shows: while \gls{cra} has no gain over \gls{maf}, \gls{pps} can select the nodes to poll more efficiently, reducing the average \gls{aoii} by about $30\%$.
Due to quantization, the performance gain of \gls{pps} is not constant in terms of the $99$th percentile performance, as shown by Fig.~\ref{fig:pull_load_99}: in this case, the gain with respect to \gls{maf} is $10$~ms, i.e., one frame, except for a few settings. On the other hand, \gls{cra} tends to perform worse than \gls{maf}. In general, \gls{pps} never performs worse than the benchmark algorithms, and performs much better in several conditions.

\subsection{Performance: DT Drift in Kalman scenario}
\label{sec:results:dt-drift-kalman}

\begin{figure}
\centering
\subfloat[Average \gls{mse} $\E{e(\mb{x}, \hat{\mb{x}})}$. \label{fig:pull_frame_kalman_avg}]{\begin{tikzpicture}
\pgfplotstableread{fig/fig_data/pull_frame_kalman_avg.csv}\loadedtable;
\begin{axis}[
    width=\fwidth,
    height=\fheight,
    ybar=0.5pt,
    tick align=inside,
    bar width=1.5pt,
    legend style={font=\scriptsize, at={(1, 1)}, anchor=north east, legend columns=3},
    xmin=4.5,
    xmax=20.5,
    xlabel={$Q$ (REs)},
    ymin=0,
    ymax=15,
    ylabel={$\E{e(\mb{x}, \hat{\mb{x}})}$},
     legend image code/.code={\draw [#1] (0cm,-0.1cm) rectangle (0.15cm,4pt); },
]

    \addplot[style={color0,fill={white!20!color0}}] table[x=Q, y=MAF] {\loadedtable}; 
    \addlegendentry{MAF};
    \addplot[style={color2,fill={white!20!color2}}] table[x=Q, y=CRA] {\loadedtable}; 
    \addlegendentry{CRA};
    \addplot[style={color4,fill={white!20!color4}}] table[x=Q, y=PPS] {\loadedtable}; 
    \addlegendentry{PPS};
\end{axis}
\end{tikzpicture}}\hfill
\subfloat[$99$th percentile \gls{mse} $e(\mb{x}, \hat{\mb{x}})_{99}$. \label{fig:pull_frame_kalman_99}]{\begin{tikzpicture}
\pgfplotstableread{fig/fig_data/pull_frame_kalman_99.csv}\loadedtable;
\begin{axis}[
    width=\fwidth,
    height=\fheight,
    ybar=0.5pt,
    tick align=inside,
    bar width=1.5pt,
    legend style={font=\tiny, at={(1, 1)}, anchor=north east, legend columns=1},
    xmin=4.5,
    xmax=20.5,
    xlabel={$Q$ (REs)},
    ymin=0,
    ymax=100,
    ylabel={$e(\mb{x}, \hat{\mb{x}})_{99}$},
     legend image code/.code={\draw [#1] (0cm,-0.1cm) rectangle (0.15cm,4pt); },
]

    \addplot[style={color0,fill={white!20!color0}}] table[x=Q, y=MAF] {\loadedtable}; 
    \addplot[style={color2,fill={white!20!color2}}] table[x=Q, y=CRA] {\loadedtable}; 
    \addplot[style={color4,fill={white!20!color4}}] table[x=Q, y=PPS] {\loadedtable}; 
\end{axis}
\end{tikzpicture}}
\caption{Average and worst-case \gls{mse} for \gls{dt} drift as a function of $Q$  under a purely pull-based system with $\sigma_w = 0.5$.}\vspace{-0.3cm}
\label{fig:pull_frame_kalman}
\end{figure}

 We now analyze the \gls{dt} alignment performance in Scenario~\ref{scen:kalman} for a purely pull-based system, i.e., $Q = R$. First, we test the behavior of the system as a function of the available \glspl{re} $Q$, considering the well-specified model in~\eqref{eq:F-well}, with $\sigma_w = 0.5$.
 Fig.~\ref{fig:pull_frame_kalman} shows the average and the worst case \gls{mse} for \gls{maf}, \gls{cra}, and the \gls{pps} approaches. While~\gls{cra} and \gls{maf} performance are comparable, \gls{pps} outperforms the two benchmark: for $5 \le Q \le 15$, it reduces the average \gls{mse} by at least $11$\% and the worst-case  \gls{mse} by at least $25$\%, with peaks of $\approx 32\%$ and $\approx60\%$, respectively, for $Q=8$. The \Gls{pps} gain is lower for $Q >15$, with a minimum for $Q=20$ of $\approx 1$\% on average and $\approx14$\% on the $99$th percentile \gls{mse}.

\begin{figure}[t]
\centering
\subfloat[Average \gls{mse} $\E{e(\mb{x}, \hat{\mb{x}})}$. \label{fig:pull_process_avg}]{\begin{tikzpicture}
\pgfplotstableread{fig/fig_data/pull_process_kalman_avg.csv}\loadedtable;
\begin{axis}[
    width=.41\fwidth,
    height=\fheight,
    xlabel={$\sigma_w$},
    ylabel={$\E{e(\mb{x}, \hat{\mb{x}})}$},
    xmin=0, xmax=1,
    ymin=0, ymax=20,    
    legend style={font=\scriptsize, at={(1,0.5)}, legend columns=1, anchor=west, draw=none, fill=none}
]

\addplot[
    color=color0, mark=triangle,mark repeat=5
    ]
    table[x=sigma_w,y=MAF] {\loadedtable};
\addlegendentry{MAF}; 
\addplot[
    color=color2, mark=diamond,mark repeat=5, mark phase=3
    ]
    table[x=sigma_w,y=CRA] {\loadedtable};
\addlegendentry{CRA};
\addplot[
    color=color4, mark=square, mark repeat=5
    ]
    table[x=sigma_w,y=PPS] {\loadedtable};
\addlegendentry{PPS};

\end{axis}
\end{tikzpicture}}
\subfloat[$99$th percentile \gls{mse} $e(\mb{x}, \hat{\mb{x}})_{99}$. \label{fig:pull_process_99}]{\begin{tikzpicture}
\pgfplotstableread{fig/fig_data/pull_process_kalman_99.csv}\loadedtable;
\begin{axis}[
    width=.5\fwidth,
    height=\fheight,
    xlabel={$\sigma_w$},
    ylabel={$e(\mb{x}, \hat{\mb{x}})_{99}$},
    xmin=0, xmax=1,
    ymin=0, ymax=100,
]

\addplot[
    color=color0, mark=triangle,mark repeat=5
    ]
    table[x=sigma_w,y=MAF] {\loadedtable};
\addplot[
    color=color2, mark=diamond,mark repeat=5, mark phase=3
    ]
    table[x=sigma_w,y=CRA] {\loadedtable};
\addplot[
    color=color4, mark=square, mark repeat=5
    ]
    table[x=sigma_w,y=PPS] {\loadedtable};

\end{axis}
\end{tikzpicture}}
\caption{Average and worst-case \gls{mse} for \gls{dt} drift as a function of the process \gls{sd} $\sigma_w$ under a purely pull-based system with $Q = 10$.}\vspace{-0.3cm}
\label{fig:pull_process}
\end{figure}

\begin{figure}[t]
\centering
\subfloat[Average \gls{mse} $\E{e(\mb{x},\hat{\mb{x}})}$. \label{fig:pull_noise_avg}]{\begin{tikzpicture}
\pgfplotstableread{fig/fig_data/pull_noise_kalman_50_avg.csv}\loadedtable;
\begin{axis}[
    width=.42\fwidth,
    height=\fheight,
    xlabel={$\hat{\sigma}_w$},
    ylabel={$\E{e(\mb{x}, \hat{\mb{x}})}$},
    xmin=0, xmax=1,
    ymin=0, ymax=10,    
    legend style={font=\scriptsize, at={(1,0.5)}, legend columns=1, anchor=west, draw=none, fill=none},    
]

\addplot[
    color=color0, mark=triangle,mark repeat=5
    ]
    table[x=sigma_w_hat,y=MAF] {\loadedtable};
\addlegendentry{MAF}; 
\addplot[
    color=color2, mark=diamond,mark repeat=5, mark phase=3
    ]
    table[x=sigma_w_hat,y=CRA] {\loadedtable};
\addlegendentry{CRA};
\addplot[
    color=color4, mark=square, mark repeat=5
    ]
    table[x=sigma_w_hat,y=PPS] {\loadedtable};
\addlegendentry{PPS};

\end{axis}
\end{tikzpicture}}
\subfloat[$99$th percentile \gls{mse} $e(\mb{x},\hat{\mb{x}})_{99}$. \label{fig:pull_noise_99}]{\begin{tikzpicture}
\pgfplotstableread{fig/fig_data/pull_noise_kalman_50_99.csv}\loadedtable;
\begin{axis}[
    width=.5\fwidth,
    height=\fheight,
    xlabel={$\hat{\sigma}_w$},
    ylabel={$e(\mb{x},\hat{\mb{x}})_{99}$},
    xmin=0, xmax=1,
    ymin=0, ymax=50,    
    legend style={font=\scriptsize, at={(0,1)}, draw=none, fill=none, legend columns=2, anchor=north west},    
]

\addplot[
    color=color0, mark=triangle,mark repeat=5
    ]
    table[x=sigma_w_hat,y=MAF] {\loadedtable};
\addplot[
    color=color2, mark=diamond,mark repeat=5, mark phase=3
    ]
    table[x=sigma_w_hat,y=CRA] {\loadedtable};
\addplot[
    color=color4, mark=square, mark repeat=5
    ]
    table[x=sigma_w_hat,y=PPS] {\loadedtable};

\end{axis}
\end{tikzpicture}}
\caption{Average and worst-case \gls{mse} for \gls{dt} drift as a function of the estimated process noise \gls{sd} $\hat{\sigma}_w$ while keeping $\sigma_w = 0.5$, under a purely pull-based system with $Q = 10$.}\vspace{-0.3cm}
\label{fig:pull_noise}
\end{figure}

Next, we set $Q=10$ and analyze \gls{dt} drift as a function of the process \gls{sd} $\sigma_w$. Fig.~\ref{fig:pull_process} shows that \gls{pps} always outperforms the benchmarks, with gains ranging within $9-25\%$ on average \gls{mse} and $23-56\%$ on worst-case performance. A substantial gain is found for $\sigma_w \le 0.75$, with at least a $42\%$ improvement on $99$th percentile \gls{mse}.
\Gls{maf} performs slightly better than \gls{cra}, but the difference is almost negligible. 

We assess the performance of \gls{pps} under mis-specified models. Fig.~\ref{fig:pull_noise} shows the impact on errors in estimating the process noise \gls{sd}, by letting drift be generated with $\sigma_w = 0.5$, while the system works with an estimate $\hat{\sigma}_w\in(0,1]$. Results show that \gls{sd} mis-estimation impacts performance only when process noise is heavily underestimated: \gls{pps} outperforms the benchmarks for $\hat{\sigma}_w \ge 0.1$, and has stable performance for $\hat{\sigma}_w \ge 0.2$.
Fig.~\ref{fig:pull_misspecified} assesses performance as a function of $\beta_F$ as defined in~\eqref{eq:F-mis}, considering mis-estimation of the dynamic system evolution. The results show that \gls{maf} is the least affected by a mis-specified transition matrix, as it only focuses on nodes' age to schedule updates. Conversely, \gls{pps} strongly relies on the prior model for node scheduling; however, it still outperforms the benchmarks for $\beta_F < 0.09$. We remark that if $\beta_F$ converges to $1/9 \simeq 0.11$, $\hat{\mb{F}}$ becomes the identity matrix, heavily impacting the estimated state update model.
Due to cluster-by-cluster scheduling, \gls{cra} has a slightly stabler performance, but starts to diverge from $\beta_F \ge 0.08$.

\begin{figure}[t]
\centering
\subfloat[Average \gls{mse} $\E{e(\mb{x}, \hat{\mb{x}})}$. \label{fig:pull_misspecified_avg}]{\begin{tikzpicture}
\pgfplotstableread{fig/fig_data/pull_misspec_kalman_50_avg.csv}\loadedtable;
\begin{axis}[
    width=0.44\fwidth,
    height=\fheight,
    xlabel={$\beta_F$},
    ylabel={$\E{e(\mb{x}, \hat{\mb{x}})}$},
    xmin=-0.11, xmax=0.11,
    ymin=0, ymax=3,    
    legend style={font=\scriptsize, at={(1,0.5)}, legend columns=1, anchor=west, draw=none, fill=none}, 
    xtick={-0.1, -0.05, 0, 0.05, 0.1},
    xticklabels={-0.1, -0.05, 0, 0.05, 0.1}
]

\addplot[
    color=color0, mark=triangle,mark repeat=5
    ]
    table[x=delta_F,y=MAF] {\loadedtable};
\addlegendentry{MAF}; 
\addplot[
    color=color2, mark=diamond,mark repeat=5, mark phase=3
    ]
    table[x=delta_F,y=CRA] {\loadedtable};
\addlegendentry{CRA};
\addplot[
    color=color4, mark=square, mark repeat=5
    ]
    table[x=delta_F,y=PPS] {\loadedtable};
\addlegendentry{PPS};

\end{axis}
\end{tikzpicture}}\hfill
\subfloat[$99$th percentile \gls{mse} $e(\mb{x}, \hat{\mb{x}})_{99}$. \label{fig:pull_misspecified_99}]{\begin{tikzpicture}
\pgfplotstableread{fig/fig_data/pull_misspec_kalman_50_99.csv}\loadedtable;
\begin{axis}[
    width=0.5\fwidth,
    height=\fheight,
    xlabel={$\beta_F$},
    ylabel={$e(\mb{x}, \hat{\mb{x}})_{99}$},
     xmin=-0.11, xmax=0.11,
    ymin=0, ymax=30,    
    legend style={font=\scriptsize, at={(0,1)}, legend columns=2, anchor=north west}, 
    xtick={-0.1, -0.05, 0, 0.05, 0.1},
    xticklabels={-0.1, -0.05, 0, 0.05, 0.1}
]

\addplot[
    color=color0, mark=triangle,mark repeat=5
    ]
    table[x=delta_F,y=MAF] {\loadedtable};
\addplot[
    color=color2, mark=diamond,mark repeat=5, mark phase=3
    ]
    table[x=delta_F,y=CRA] {\loadedtable};
\addplot[
    color=color4, mark=square, mark repeat=5
    ]
    table[x=delta_F,y=PPS] {\loadedtable};

\end{axis}
\end{tikzpicture}}
\caption{Average and worst-case \gls{mse} for \gls{dt} drift as a function of $\beta_F$, modeling an error between the ground-truth $\mb{F}$~\eqref{eq:F-well} and its estimated counterpart $\hat{\mb{F}}$~\eqref{eq:F-mis}, under a purely pull-based system with $Q = 10$ and $\hat{\sigma}_w = \sigma_w = 0.5$.}\vspace{-0.3cm}
\label{fig:pull_misspecified}
\end{figure}

\subsection{Performance: Anomaly Tracking}

\begin{figure}
\centering
\subfloat[Average \gls{aoii} $\E{\Theta}$. \label{fig:push_load_avg}]{\begin{tikzpicture}
\begin{axis}[
    width=.42\fwidth,
    height=\fheight,
    xlabel={$\rho_a$ (Hz)},
    ylabel={$\E{\Theta}$ (ms)},
    xmin=1, xmax=5,
    ymin=0, ymax=3,
    legend pos=south east,
    ytick={0,0.5,1,1.5,2,2.5,3},
    yticklabels={0,5,10,15,20,25,30},
    legend style={font=\scriptsize, at={(1,0.5)}, legend columns=1, anchor=west, draw=none, fill=none}
]

\addplot[
    color=color0, mark=triangle,mark repeat=5
    ]
    table[x=rate,y=RR] {fig/fig_data/push_load_avg.csv};
\addlegendentry{MAF};
\addplot[
    color=color1, mark=o,mark repeat=5
    ]
    table[x=rate,y=aloha] {fig/fig_data/push_load_avg.csv};
\addlegendentry{FSA}; 
\addplot[
    color=color3, mark=x,mark repeat=5
    ]
    table[x=rate,y=adapt] {fig/fig_data/push_load_avg.csv};
\addlegendentry{AFSA};
\addplot[
    color=color4, mark=square,mark repeat=5
    ]
    table[x=rate,y=PPS] {fig/fig_data/push_load_avg.csv};
\addlegendentry{PPS};

\end{axis}
\end{tikzpicture}}~
\subfloat[$99$th percentile \gls{aoii} $\Theta_{99}$. \label{fig:push_load_99}]{\begin{tikzpicture}
\begin{axis}[
    width=.49\fwidth,
    height=\fheight,
    xlabel={$\rho_a$ (Hz)},
ylabel={$\Theta_{99}$ (ms)},
    xmin=1, xmax=5,
    ymin=0, ymax=20,
    ytick={0,5,10,15,20,25,30},
    yticklabels={0,50,100,150,200,250,300},
    legend style={legend columns=1, font=\scriptsize, anchor=north west, at={(0.01,0.98)}}
]

\addplot[
    color=color0, mark=triangle,mark repeat=5
    ]
    table[x=rate,y=RR] {fig/fig_data/push_load_99.csv};
\addplot[
    color=color1, mark=o,mark repeat=5
    ]
    table[x=rate,y=aloha] {fig/fig_data/push_load_99.csv};
\addplot[
    color=color3, mark=x,mark repeat=5
    ]
    table[x=rate,y=adapt] {fig/fig_data/push_load_99.csv};
\addplot[
    color=color4, mark=square,mark repeat=5
    ]
    table[x=rate,y=PPS] {fig/fig_data/push_load_99.csv};

\end{axis}
\end{tikzpicture}}
\caption{Average and worst-case anomaly \gls{aoii} as a function of $\rho_a$ under a purely push-based system. with $P=10$.}
\label{fig:push_load}
\end{figure}

\begin{figure}[t]
\centering
\subfloat[Average \gls{aoii} $\E{\Theta}$. \label{fig:push_frame_avg}]{\begin{tikzpicture}
\pgfplotstableread{fig/fig_data/push_frame_sota_avg.csv}\loadedtable;
\begin{axis}[
    width=\fwidth,
    height=\fheight,
    ybar=0.5pt,
    tick align=inside,
    bar width=1.5pt,
    legend style={font=\scriptsize, at={(1,1)}, legend columns=2, anchor=north east},
    xmin=4.5,
    xmax=20.5,
    xlabel={$P$ (REs)},
    ymin=0,
    ymax=5,
    ytick={0,1,2,3,4,5},
    yticklabels={0,10,20,30,40,50},
    ylabel={$\E{\Theta}$ (ms)},    
    legend image code/.code={\draw [#1] (0cm,-0.1cm) rectangle (0.15cm,4pt); },
]
    \addplot[style={color0,fill={white!20!color0}}] table[x=P, y=MAF] {\loadedtable};
    \addlegendentry{MAF};
    \addplot[style={color1,fill={white!20!color1}}] table[x=P, y=FSA] {\loadedtable}; 
    \addlegendentry{FSA};
    \addplot[style={color3,fill={white!20!color3}}] table[x=P, y=AFSA] {\loadedtable}; 
    \addlegendentry{AFSA};
    \addplot[style={color4,fill={white!20!color4}}] table[x=P, y=PPS] {\loadedtable}; 
    \addlegendentry{PPS};
\end{axis}
\end{tikzpicture}}\\
\subfloat[$99$th percentile \gls{aoii} $\Theta_{99}$.\label{fig:push_frame_99}]{\begin{tikzpicture}
\pgfplotstableread{fig/fig_data/push_frame_sota_99.csv}\loadedtable;
\begin{axis}[
    width=\fwidth,
    height=\fheight,
    ybar=0.5pt,
    tick align=inside,    
    bar width=1.5pt,
    xmin=4.5,
    xmax=20.5,
    xlabel={$P$ (REs)},
    ymin=0,
    ymax=25,
    ytick={0,5,10,15,20,25,30},
    yticklabels={0,50,100,150,200,250,300},
    ylabel={$\Theta_{99}$ (ms)},
]
    \addplot[style={color0,fill={white!20!color0}}] table[x=P, y=MAF] {\loadedtable};
    \addplot[style={color1,fill={white!20!color1}}] table[x=P, y=FSA] {\loadedtable}; 
    \addplot[style={color3,fill={white!20!color3}}] table[x=P, y=AFSA] {\loadedtable}; 
    \addplot[style={color4,fill={white!20!color4}}] table[x=P, y=PPS] {\loadedtable}; 
\end{axis}
\end{tikzpicture}}
\caption{Average and worst-case anomaly \gls{aoii} as a function of $P$ under a purely push-based system.}
\label{fig:push_frame}
\end{figure}

We evaluate the anomaly tracking performance of \gls{pps}.
Fig.~\ref{fig:push_load} shows the performance of \gls{pps} as a function of the anomaly rate $\rho_a$, in a scenario in which we only consider the push subframes (i.e., $R = P$) and $P=10$. If the load is very low, random access mechanisms such as \gls{fsa} are able to perform well, since collisions are extremely rare. We note that \gls{pps} performs as well as \gls{fsa} in these conditions, for both the average and $99$th percentile \gls{aoii}, $\Theta_{99}$. However, \gls{fsa} is highly unstable at higher loads, while the performance of \gls{pps} degrades much more gracefully as the load increases. While \gls{maf} remains the best solution for very high rates, \gls{pps} maintains an advantage over both the expected \gls{aoii} and worst-case performance for anomaly rates up to $4$~Hz.

\begin{figure}[t]
\centering
\subfloat[Average. \label{fig:coexistence_frame_hom_avg}]{\begin{tikzpicture}
\pgfplotstableread{fig/fig_data/coexistence_frame_hom.csv}\tablehom;
\pgfplotstableread{fig/fig_data/coexistence_frame_het.csv}\tablehet;
\begin{axis}[
    width=\fwidth,
    height=\fheight,
    ybar=0.5pt,
    tick align=inside,
    bar width=2pt,
    legend style={font=\scriptsize, at={(0.35, 1)}, anchor=north, legend columns=1},
    xmin=2.5,
    xmax=17.5,
    xlabel={$P$ (REs)},
    ymin=0,
    ymax=1.5,
    ytick={0,0.25,0.5,0.75,1,...,3},
    yticklabels={0,2.5,5,7.5,10,...,30},
    ylabel={$\E{x}$ (ms)},
    legend image code/.code={\draw [#1] (0cm,-0.1cm) rectangle (0.15cm,4pt); },
]

    \addplot[style={color0,fill={white!20!color0}}] table[x=P, y=PsiAvg] {\tablehom}; 
    \addlegendentry{$\Psi$, $\rho_m\!=\!\rho_a\!=\!3$~Hz}
    \addplot[style={color3,fill={white!20!color3}}] table[x=P, y=ThetaAvg] {\tablehom};     
    \addlegendentry{$\Theta$, $\rho_m\!=\!\rho_a\!=\!3$~Hz};    
    \addplot[color0,fill={white!20!color0},
    postaction={pattern=north east lines}] table[x=P, y=PsiAvg] {\tablehet}; 
    \addlegendentry{$\Psi$, $\rho_m\!=\!2.5$~Hz, $\rho_a\!=\!3.5$~Hz}
    \addplot[color3,fill={white!20!color3},
    postaction={pattern=north east lines}] table[x=P, y=ThetaAvg] {\tablehet}; 
    \addlegendentry{$\Theta$, $\rho_m\!=\!2.5$~Hz, $\rho_a\!=\!3.5$~Hz};    
\end{axis}
\end{tikzpicture}}\\ 
\subfloat[$99$th percentile. \label{fig:coexistence_frame_hom_99}]{\begin{tikzpicture}
\pgfplotstableread{fig/fig_data/coexistence_frame_hom.csv}\tablehom;
\pgfplotstableread{fig/fig_data/coexistence_frame_het.csv}\tablehet;
\begin{axis}[
    width=\fwidth,
    height=\fheight,
    ybar=0.5pt,
    tick align=inside,
    bar width=2pt,
    legend style={font=\tiny, at={(0.5, 1)}, anchor=north, legend columns=2},
    xmin=2.5,
    xmax=17.5,
    xlabel={$P$ (REs)},
    ymin=0,
    ymax=15,
    ytick={0,2,4,...,30},
    yticklabels={0,20,40,...,300},
    ylabel={$x_{99}$ (ms)},
]

    \addplot[style={color0,fill={white!20!color0}}] table[x=P, y=Psi99] {\tablehom}; 
    \addplot[style={color3,fill={white!20!color2}}] table[x=P, y=Theta99] {\tablehom}; 
    \addplot[color0,fill={white!20!color0},
    postaction={pattern=north east lines}] table[x=P, y=Psi99] {\tablehet}; 
    \addplot[color3,fill={white!20!color3},
    postaction={pattern=north east lines}] table[x=P, y=Theta99] {\tablehet}; 
\end{axis}
\end{tikzpicture}} 
\caption{Average and worst-case \gls{aoii} for \gls{dt} drift (blue) and anomalies (red) as a function of $P$ using \gls{pps} for pull and push subframes (Scenario~\ref{scen:hmm}).}
\label{fig:coexistence_frame}
\end{figure}

We further analyze the schemes performance by considering the anomaly \gls{aoii} for different values of $P$, shown in Fig.~\ref{fig:push_frame} for a scenario with $\rho_a=3$~Hz: a higher number of \glspl{re} in the push subframe reduces the overall load on the system, making the scenario easier for the medium access schemes. Fig.~\ref{fig:push_frame_avg} shows the average performance, and we can immediately notice that \gls{pps} is the best-performing algorithm for $P\geq7$, with a much better performance than the other random access algorithms for scarce wireless resources. Its average \gls{aoii} is lower than $1$~ms with $P\geq10$, while \gls{fsa} reaches that threshold only with $P\geq12$, using $20\%$ more resources. On the other hand, \gls{maf} performs better than \gls{pps} when $P\leq6$, but the gains from additional resources are very limited, as it is limited by its scheduling-based nature, which cannot consider the actual anomalies observed by nodes.
The limited gains are also noticeable in Fig.~\ref{fig:push_frame_99}, which shows the $99$th percentile of the \gls{aoii} of anomalies: while \gls{pps} manages to quickly reduce the worst-case \gls{aoii} with more available resources, \gls{maf} has an unimpressive worst-case performance when there are plenty of transmission opportunities. 
\Gls{fsa} has a good performance when $P\geq12$, but quickly degrades when fewer resources are assigned to the push subframe, while \gls{afsa} degrades more gracefully, but is still significantly worse than \gls{pps}.

\subsection{Performance: Joint DT Alignment and Anomaly Tracking}
In this section, we evaluate the joint performance of \gls{dt} alignment and anomaly tracking, considering the full push-pull system in the two scenarios under consideration.

First, we assess \gls{pps} performance 
when using a fixed \gls{re} allocation, i.e, with $P$ \glspl{re} assigned to the push subframe and $R - P = Q$ to the pull subframe.
Fig.~\ref{fig:coexistence_frame} shows the \gls{dt} drift and anomaly \gls{aoii} performance in Scenario~\ref{scen:hmm}. The system is tested under homogeneous rates, i.e., $\rho_d = \rho_a = 3$~Hz, and heterogeneous rates, i.e., $\rho_d = 2.5$~Hz and $\rho_a = 3.5$~Hz. Note that an imbalance in the rates skews the \gls{aoii} performance, which degrades for anomalies and improves for \gls{dt} drift. This skew is more prominent in the average than for the $99$th percentile.
In both cases, performance improves with respect to what can be achieved by systems using only the push- or pull-based paradigms: this is clear when comparing the results for $P=10$ with the ones in Fig.~\ref{fig:pull_load} and Fig.~\ref{fig:push_load}.
Also, Fig.~\ref{fig:cdf} shows the \gls{cdf} of the \gls{dt} drift \gls{mse} and anomaly \gls{aoii} obtained in Scenario~\ref{scen:kalman}. Again, combining push and pull subframes improves performance: considering $P = Q=10$, we have $e(\mb{x}, \hat{\mb{x}})_{99} = 15.45$ against $17.70$ of the pull-only case (see Fig.~\ref{fig:pull_frame_kalman}) and $\Theta_{99} = 10$~ms against $30$~ms for the push-only setting (see Fig.~\ref{fig:push_frame}).
The previous effect is due to the \emph{interdependence} between the \gls{dt} drift minimization and anomaly detection tasks in our system: as explained in Sec.~\ref{sec:tracking}, data received during the pull subframe are also used to detect anomalies, and vice versa.

\begin{figure}[t]
    \centering
    \subfloat[\gls{dt} drift MSE \label{fig:cdf_mse}]{\begin{tikzpicture}

\pgfplotstableread{fig/fig_data/coexistence_kalman_cdf_mse.csv}\table;

\begin{axis}[
    width=0.5\fwidth,
    height=\fheight,
    xlabel={$e(\mb{x},\hat{\mb{x}})$},
    ylabel={CDF},
    xmin=0, xmax=40,
    ymin=0, ymax=1,
    legend style={legend columns=1, font=\scriptsize, anchor=south east, at={(1,0)}}
]

\addplot[
    color=color0, mark=triangle, mark repeat=25
    ]
    table[x=MSE,y=05]{\table};
\addlegendentry{$P = 5$};
\addplot[
    color=color1, mark=o,mark repeat=25
    ]
    table[x=MSE,y=10] {\table};
\addlegendentry{$P = 10$};
\addplot[
    color=color3, mark=x,mark repeat=25
    ]
    table[x=MSE,y=15] {\table};
\addlegendentry{$P = 15$};

\end{axis}
\end{tikzpicture}}
    \hfill
    \subfloat[Anomaly \gls{aoii}\label{fig:cdf_aoii}]{\begin{tikzpicture}

\pgfplotstableread{fig/fig_data/coexistence_kalman_cdf_aoii.csv}\table;

\begin{axis}[
    width=0.5\fwidth,
    height=\fheight,
    xlabel={$\Theta$ (ms)},
    ylabel={CDF},
    xmin=0, xmax=5,
    xtick={0,1,2,3,4,5},
    xticklabels={0,10,20,30,40,50},
    ytick={0.95,0.96,0.97,0.98,0.99,1},
    ymin=0.95, ymax=1,
    legend style={legend columns=1, font=\scriptsize, anchor=north west, at={(0.01,0.98)}}
]

\addplot+[const plot,
    color=color0, mark=triangle
    ]
    table[x=AoII,y=05]{\table};
\addplot+[const plot,
    color=color1, mark=o
    ]
    table[x=AoII,y=10] {\table};
\addplot+[const plot,
    color=color3, mark=x
    ]
    table[x=AoII,y=15] {\table};

\end{axis}
\end{tikzpicture}}
    \caption{\gls{cdf} of \gls{dt} drift MSE $e(\mb{x},\hat{\mb{x}})$ and anomaly \gls{aoii} $\Theta$ for different values of $P$ achieved by \gls{pps} (Scenario~\ref{scen:kalman}).}
    \label{fig:cdf}
\end{figure}

\begin{figure}
     \centering
    \subfloat[$\min_{P} \E{\Psi}$ s.t. $\Theta_{99} \le 30$ ms. \label{fig:minPsi}]{\begin{tikzpicture}

\begin{axis}[
    width=\fwidth,
    height=\fheight,
    ybar=1.5pt,
    ylabel={$\E{\Psi}$ (ms)},
    ymin=0, 
    ymax=1.5,
    xtick={0, 1,...,13},    
    xticklabels={MAF-MAF, MAF-FSA, MAF-AFSA,  CRA-MAF, CRA-FSA, CRA-AFSA, MAF-\textbf{PPS}, CRA-\textbf{PPS}, \textbf{PPS}-MAF, \textbf{PPS}-FSA, \textbf{PPS}-AFSA, \textbf{PPS}-\textbf{PPS}, RSM, SSM},
    xmin=-0.5,
    xmax=13.5,
    legend style={font=\scriptsize, at={(0.5, 1)}, anchor=north, legend columns=1},    
    xticklabel style={xshift=3pt,yshift=-2pt,font=\scriptsize,rotate=40,anchor=east}, 
    tick align=inside,
]

\addplot [style={color0,fill={white!20!color0}}, bar shift=0pt] 
coordinates {(0, 1.32341)}; 
\addplot[style={color0,fill={white!20!color1}}, bar shift=0pt] coordinates {(1, 1.33458)}; 
\addplot[style={color0,fill={white!20!color3}}, bar shift=0pt] coordinates {(2, 0)}; 
\addplot[style={color2,fill={white!20!color0}}, bar shift=0pt] coordinates {(3, 1.22014)}; 
\addplot[style={color2,fill={white!20!color1}}, bar shift=0pt] coordinates {(4, 1.33458)}; 
\addplot[style={color2,fill={white!20!color3}}, bar shift=0pt] coordinates {(5, 1.33569)}; 

\addplot[style={color0,fill={white!20!color4}}, bar shift=0pt] coordinates {(6, 1.13008)}; 
\addplot[style={color2,fill={white!20!color4}}, bar shift=0pt] coordinates {(7, 0.99129)}; 
\addplot[style={color4,fill={white!20!color0}}, bar shift=0pt] coordinates {(8, 1.16807)}; 
\addplot[style={color4,fill={white!20!color1}}, bar shift=0pt] coordinates {(9, 1.14704)}; 
\addplot[style={color0,fill={white!20!color3}}, bar shift=0pt] coordinates {(10, 0)}; 
\addplot[style={color4,fill={white!20!color4}}, bar shift=0pt] coordinates {(11, 0.96801)}; 
\addplot[style={color4,fill={white!20!color4}}, bar shift=0pt] coordinates {(12, 0.98839)}; 
\addplot[style={color4,fill={white!20!color4}}, bar shift=0pt] coordinates {(13, 0.96138)}; 

\node[red] at (axis cs:2, 0.5) {$\times$};
\node[red] at (axis cs:10, 0.5) {$\times$};

\draw [black, dashdotted] (axis cs:11.5, 0) -- (axis cs:11.5, 1.5);
\coordinate (overflowmark1) at (axis cs:5,1.5);
\coordinate (overflowmark2) at (axis cs:12.5,1.5);
\end{axis}

\node[above, font=\footnotesize] at (overflowmark1){Fixed $P/Q$};    
\node[above, font=\footnotesize] at (overflowmark2){Adaptive};    
\end{tikzpicture}}\\
    \subfloat[$\min_{P} \Theta_{99}$ s.t. $\E{\Psi} \le 1$ ms. \label{fig:minTheta}]{\begin{tikzpicture}

\begin{axis}[
    width=\fwidth,
    height=\fheight,
    ybar=1.5pt,
    ylabel={$\Theta_{99}$ (ms)},
    ymin=0, 
    ymax=100,
    xtick={0, 1,...,13},    
    xticklabels={MAF-MAF, MAF-FSA, MAF-AFSA, CRA-MAF, CRA-FSA, CRA-AFSA, MAF-\textbf{PPS}, CRA-\textbf{PPS}, \textbf{PPS}-MAF, \textbf{PPS}-FSA, \textbf{PPS}-AFSA, \textbf{PPS}-\textbf{PPS}, RSM, SSM},
    xmin=-0.5,
    xmax=13.5,
    legend style={font=\scriptsize, at={(0.5, 1)}, anchor=north, legend columns=1},     
    xticklabel style={xshift=3pt,yshift=-2pt,font=\scriptsize,rotate=40,anchor=east}, 
    tick align=inside,
]

\addplot [style={color0,fill={white!20!color0}}, bar shift=0pt] 
coordinates {(0, 70)}; 
\addplot[style={color0,fill={white!20!color1}}, bar shift=0pt] coordinates {(1, 650)}; 
\addplot[style={color0,fill={white!20!color3}}, bar shift=0pt] coordinates {(2, 690)}; 
\addplot[style={color2,fill={white!20!color0}}, bar shift=0pt] coordinates {(3, 80)}; 
\addplot[style={color2,fill={white!20!color1}}, bar shift=0pt] coordinates {(4, 630)}; 
\addplot[style={color2,fill={white!20!color3}}, bar shift=0pt] coordinates {(5, 590)}; 
\addplot[style={color0,fill={white!20!color4}}, bar shift=0pt] coordinates {(6, 590)};  
\addplot[style={color2,fill={white!20!color4}}, bar shift=0pt] coordinates {(7, 30)};  
\addplot[style={color4,fill={white!20!color0}}, bar shift=0pt] coordinates {(8, 40)}; 
\addplot[style={color4,fill={white!20!color1}}, bar shift=0pt] coordinates {(9, 700)}; 
\addplot[style={color0,fill={white!20!color3}}, bar shift=0pt] coordinates {(10, 180)}; 
\addplot[style={color4,fill={white!20!color4}}, bar shift=0pt] coordinates {(11, 30)};  
\addplot[style={color4,fill={white!20!color4}}, bar shift=0pt] coordinates {(12, 20)}; 
\addplot[style={color4,fill={white!20!color4}}, bar shift=0pt] coordinates {(13, 30)}; 

\node[below, yshift=-2pt] at (axis cs:1,100) {$\uparrow$};
\node[below, yshift=-2pt] at (axis cs:2,100) {$\uparrow$};
\node[below, yshift=-2pt] at (axis cs:4,100) {$\uparrow$};
\node[below, yshift=-2pt] at (axis cs:5,100) {$\uparrow$};
\node[below, yshift=-2pt] at (axis cs:6,100) {$\uparrow$};
\node[below, yshift=-2pt] at (axis cs:9,100) {$\uparrow$};
\node[below, yshift=-2pt] at (axis cs:10,100) {$\uparrow$};

\draw [black, dashdotted] (axis cs:11.5, 0) -- (axis cs:11.5, 100);
\coordinate (overflowmark1) at (axis cs:5,100);
\coordinate (overflowmark2) at (axis cs:12.5,100);
\end{axis}

\node[above, font=\footnotesize] at (overflowmark1){Fixed $P/Q$};    
\node[above, font=\footnotesize] at (overflowmark2){Adaptive}; 
\end{tikzpicture}}    
    \caption{Best achievable performance in the \gls{hmm} scenario when constraining the push-pull system to target a specific metric $\E{\Psi}$ or $\Theta_{99}$. The \textcolor{red}{$\times$} and the $\uparrow$ represent that no value satisfy the requirements or is out of the limit of the axis, respectively, and $\rho_d = \rho_a = 3$~Hz.}
    \label{fig:managers-discrete}
\end{figure}

Finally, we evaluate the dual optimization of \gls{dt} drift and anomaly tracking by setting a performance constraint for one task and evaluating the other. 
Figs.~\ref{fig:managers-discrete} and~\ref{fig:managers-mixed} compare the proposed adaptive resource allocation methods to \emph{all possible combinations of schemes} for the push- and pull subframes for Scenarios~\ref{scen:hmm} and~\ref{scen:kalman}, respectively. E.g., label \gls{maf}-\gls{fsa} denotes \gls{maf} for the former and \gls{fsa} for the latter. 
The \gls{rsm} and \gls{ssm} allocation schemes employ \gls{pps} for both subframes. All results are shown for the best values of $P$ and $Q$, obtained by exhaustive search, except for \gls{rsm} and \gls{ssm}, which dynamically allocate the \glspl{re} on a frame basis.

We start by considering Scenario~\ref{scen:hmm}, for $\rho_d = \rho_a = 3$~Hz. 
Fig.~\ref{fig:minPsi} shows the average \gls{dt} drift \gls{aoii} when setting the $99$th percentile of the anomaly \gls{aoii} to be at most $30$~ms. The choice of using worst-case performance for anomalies and average performance for \gls{dt} drift is due to the more critical nature of the former, which need to be reliably reported, while the latter represents the standard operation of the system, and should thus be optimized in the average sense. \gls{pps}-based schemes have a clear advantage, as they are the only ones to achieve $\E{\Psi}\leq1$~ms, gaining approximately $20\%$ over \gls{cra}-\gls{maf}, the best combination of state-of-the-art schemes. In this case, \gls{rsm} and \gls{ssm} achieve approximately the same performance as the best static \gls{pps} configuration, without the need to find the best \gls{re} configuration beforehand, which might be prohibitive for fluctuating system conditions.
Fig.~\ref{fig:minTheta} considers the dual problem, setting $\E{\Psi}\leq1$~ms as a system constraint: in this case, the performance gain of \gls{pps} is much more evident, as \gls{rsm} is able to achieve $\Theta_{99}=20$~ms, while state-of-the-art combinations all have at least $70$~ms. Interestingly, \gls{cra}-\gls{pps} performs almost as well as pure \gls{pps}, followed by \gls{pps}-\gls{maf}, showing that using \gls{pps} for just one of the two subframes may significantly improve performance.

\begin{figure}
     \centering
    \subfloat[$\min_{P} \E{e(\mb{x}, \hat{\mb{x}})}$ s.t. $\Theta_{99} \le 30$ ms. \label{fig:minPsi_kalman}]{\begin{tikzpicture}

\begin{axis}[
    width=\fwidth,
    height=\fheight,
    ybar=1.5pt,
    ylabel={$\E{\Psi}$ (ms)},
    ymin=0, 
    ymax=3.,
    xtick={0, 1,...,13},    
    xticklabels={MAF-MAF, MAF-FSA, MAF-AFSA,  CRA-MAF, CRA-FSA, CRA-AFSA, MAF-\textbf{PPS}, CRA-\textbf{PPS}, \textbf{PPS}-MAF, \textbf{PPS}-FSA, \textbf{PPS}-AFSA, \textbf{PPS}-\textbf{PPS}, RSM, SSM},
    xmin=-0.5,
    xmax=13.5,
    legend style={font=\scriptsize, at={(0.5, 1)}, anchor=north, legend columns=1},    
    xticklabel style={xshift=3pt,yshift=-2pt,font=\scriptsize,rotate=40,anchor=east}, 
    tick align=inside,
]

\addplot [style={color0,fill={white!20!color0}}, bar shift=0pt] 
coordinates {(0, 2.553296)}; 
\addplot[style={color0,fill={white!20!color1}}, bar shift=0pt] coordinates {(1, 2.50626)}; 
\addplot[style={color0,fill={white!20!color3}}, bar shift=0pt] coordinates {(2, 0)}; 
\addplot[style={color2,fill={white!20!color0}}, bar shift=0pt] coordinates {(3, 2.368151)}; 
\addplot[style={color2,fill={white!20!color1}}, bar shift=0pt] coordinates {(4, 2.797299)}; 
\addplot[style={color2,fill={white!20!color3}}, bar shift=0pt] coordinates {(5, 0)}; 

\addplot[style={color0,fill={white!20!color4}}, bar shift=0pt] coordinates {(6, 2.292925)}; 
\addplot[style={color2,fill={white!20!color4}}, bar shift=0pt] coordinates {(7, 2.290915)}; 
\addplot[style={color4,fill={white!20!color0}}, bar shift=0pt] coordinates {(8, 2.363508)}; 
\addplot[style={color4,fill={white!20!color1}}, bar shift=0pt] coordinates {(9, 2.262554)}; 
\addplot[style={color0,fill={white!20!color3}}, bar shift=0pt] coordinates {(10, 0)}; 
\addplot[style={color4,fill={white!20!color4}}, bar shift=0pt] coordinates {(11, 2.120078)}; 
\addplot[style={color4,fill={white!20!color4}}, bar shift=0pt] coordinates {(12, 2.248716)}; 
\addplot[style={color4,fill={white!20!color4}}, bar shift=0pt] coordinates {(13, 2.244635)}; 

\node[red] at (axis cs:2, 1.5) {$\times$};
\node[red] at (axis cs:5, 1.5) {$\times$};
\node[red] at (axis cs:10, 1.5) {$\times$};

\draw [black, dashdotted] (axis cs:11.5, 0) -- (axis cs:11.5, 3);
\coordinate (overflowmark1) at (axis cs:5,3);
\coordinate (overflowmark2) at (axis cs:12.5,3);
\end{axis}

\node[above, font=\footnotesize] at (overflowmark1){Fixed $P/Q$};    
\node[above, font=\footnotesize] at (overflowmark2){Adaptive};    
\end{tikzpicture}}\\
    \subfloat[$\min_{P} \Theta_{99}$ s.t. $\E{e(\mb{x}, \hat{\mb{x}})} \le 2.5$. \label{fig:minTheta_kalman}]{\begin{tikzpicture}

\begin{axis}[
    width=\fwidth,
    height=\fheight,
    ybar=1.5pt,
    ylabel={$\Theta_{99}$ (ms)},
    ymin=0, 
    ymax=100,
    xtick={0, 1,...,13},    
    xticklabels={MAF-MAF, MAF-FSA, MAF-AFSA, CRA-MAF, CRA-FSA, CRA-AFSA, MAF-\textbf{PPS}, CRA-\textbf{PPS}, \textbf{PPS}-MAF, \textbf{PPS}-FSA, \textbf{PPS}-AFSA, \textbf{PPS}-\textbf{PPS}, RSM, SSM},
    xmin=-0.5,
    xmax=13.5,
    legend style={font=\scriptsize, at={(0.5, 1)}, anchor=north, legend columns=1},     
    xticklabel style={xshift=3pt,yshift=-2pt,font=\scriptsize,rotate=40,anchor=east}, 
    tick align=inside,
]

\addplot [style={color0,fill={white!20!color0}}, bar shift=0pt] 
coordinates {(0, 50)}; 
\addplot[style={color0,fill={white!20!color1}}, bar shift=0pt] coordinates {(1, 30)}; 
\addplot[style={color0,fill={white!20!color3}}, bar shift=0pt] coordinates {(2, 160)}; 
\addplot[style={color2,fill={white!20!color0}}, bar shift=0pt] coordinates {(3, 30)}; 
\addplot[style={color2,fill={white!20!color1}}, bar shift=0pt] coordinates {(4, 770)}; 
\addplot[style={color2,fill={white!20!color3}}, bar shift=0pt] coordinates {(5, 200)}; 
\addplot[style={color0,fill={white!20!color4}}, bar shift=0pt] coordinates {(6, 20)};  
\addplot[style={color2,fill={white!20!color4}}, bar shift=0pt] coordinates {(7, 20)};  
\addplot[style={color4,fill={white!20!color0}}, bar shift=0pt] coordinates {(8, 30)}; 
\addplot[style={color4,fill={white!20!color1}}, bar shift=0pt] coordinates {(9, 20)}; 
\addplot[style={color0,fill={white!20!color3}}, bar shift=0pt] coordinates {(10, 70)}; 
\addplot[style={color4,fill={white!20!color4}}, bar shift=0pt] coordinates {(11, 20)};  
\addplot[style={color4,fill={white!20!color4}}, bar shift=0pt] coordinates {(12, 20)}; 
\addplot[style={color4,fill={white!20!color4}}, bar shift=0pt] coordinates {(13, 20)}; 

\node[below, yshift=-2pt] at (axis cs:2,100) {$\uparrow$};
\node[below, yshift=-2pt] at (axis cs:4,100) {$\uparrow$};
\node[below, yshift=-2pt] at (axis cs:5,100) {$\uparrow$};

\draw [black, dashdotted] (axis cs:11.5, 0) -- (axis cs:11.5, 100);
\coordinate (overflowmark1) at (axis cs:5,100);
\coordinate (overflowmark2) at (axis cs:12.5,100);
\end{axis}

\node[above, font=\footnotesize] at (overflowmark1){Fixed $P/Q$};    
\node[above, font=\footnotesize] at (overflowmark2){Adaptive}; 
\end{tikzpicture}}    
    \caption{Best achievable performance in the Kalman scenario when constraining the push-pull system to target a specific metric $\E{e(\mb{x}, \hat{\mb{x}})}$ or $\Theta_{99}$. The \textcolor{red}{$\times$} and the $\uparrow$ represent that no value satisfy the requirements or is out of the limit of the axis, respectively, and $\sigma_w=0.5$, $\rho_a = 3$~Hz.}
    \label{fig:managers-mixed}
\end{figure}

Then, we assess performance in Scenario~\ref{scen:kalman} with $\sigma_w = 0.5$ and $\rho_a = 3$ Hz. Considering the \gls{mse} performance, shown in Fig.~\ref{fig:minPsi_kalman}, the gain obtained by \gls{pps} when ensuring a maximum anomaly \gls{aoii} of $30$~ms is slightly smaller than in Scenario~\ref{scen:hmm}, but \gls{pps} still gains about $10$\% over \gls{cra}-\gls{maf}, the best combination of benchmark algorithms. The two dynamic managers perform slightly worse than the best \gls{pps}, but still overcome all benchmarks without any exhaustive search on the optimal \gls{re} split. The same general trend holds for the anomaly \gls{aoii}, shown in Fig.~\ref{fig:minTheta_kalman}: using \gls{pps}, or any of the two managers, ensures that the average \gls{mse} is below $2.5$, while allowing the \gls{bs} to reduce the $99$th percentile \gls{aoii} from $30$~ms (with \gls{maf}-\gls{fsa} or \gls{cra}-\gls{maf}) to $20$~ms, a gain that is smaller than in Scenario~\ref{scen:hmm}, but still significant for \gls{dt} operations. 

\section{Conclusions}\label{sec:concl}

In this work, we presented a framework that jointly optimizes \gls{dt} updates and communication system decisions and is centered on \gls{pps}, a scheduler that can exploit frame subdivision to optimize push- and pull-based traffic in a goal-oriented fashion. We considered a dual objective, targeting a system in which the \gls{dt} alignment task is complicated by the presence of anomalies that can only be detected by the sensors, and presented two heuristics that can significantly improve performance for the two tasks. We proposed dynamic resource allocation schemes, showing that an appropriate management of the pull and push subframes can lead to gain between $20\%$ and $70\%$ with respect to the state-of-the-art.

This is a first step towards the integration of push- and pull-based schemes, and future extensions will need to consider more realistic settings, as well as more complex learning-based resource allocation and scheduling schemes, to meet this challenge. Additionally, a more complete theoretical characterization of the joint problem, providing optimality guarantees and bounds, will shed further light on future protocol design.

\bibliographystyle{IEEEtran}
\bibliography{bibliography}

\end{document}